\documentclass{ar-1col}

\usepackage{graphicx}
\usepackage{amssymb}
\usepackage{amsmath}
\usepackage{natbib}
\def\gtwid{\mathrel{\raise.3ex\hbox{$>$\kern-.75em\lower1ex\hbox{$\sim$}}}}
\def\ltwid{\mathrel{\raise.3ex\hbox{$<$\kern-.75em\lower1ex\hbox{$\sim$}}}}

\def\agt{\mathrel{\raise.3ex\hbox{$>$\kern-.75em\lower1ex\hbox{$\sim$}}}}
\def\alt{\mathrel{\raise.3ex\hbox{$<$\kern-.75em\lower1ex\hbox{$\sim$}}}}

\newcommand{\be}{\begin{equation}}
\newcommand{\ee}{\end{equation}}
\newcommand{\ba}{\begin{eqnarray}}
\newcommand{\ea}{\end{eqnarray}}
\newcommand{\bu}{\boldsymbol{u}}

\newcommand{\bea}{\begin{eqnarray}}
\newcommand{\eea}{\end{eqnarray}}
\newcommand{\bean}{\begin{eqnarray}}
\newcommand{\eean}{\end{eqnarray}}

\begin{document}

\title{High Reynolds number Taylor-Couette turbulence}
\author{Siegfried  Grossmann$^1$, Detlef Lohse$^{2,3}$,  \\ and Chao Sun$^{2,4}$
\affil{$^1$ Fachbereich Physik, University of Marburg, Renthof 6, D-35032 Marburg, Germany.}
\affil{$^2$Physics of Fluids Group, Faculty  of
Science and Technology, J.M. Burgers Center for Fluid Dynamics,
and MESA+ Institute, University of Twente, The Netherlands. email addresses: d.lohse@utwente.nl; c.sun@utwente.nl.}
\affil{$^3$Max Planck Institute for Dynamics and Self-Organization, 37077 G$\ddot{o}$ttingen, Germany}
\affil{$^4$Center for Combustion Energy and Department of Thermal Engineering, Tsinghua
University, 100084 Beijing, China}}

\begin{abstract}

Taylor-Couette flow -- the flow between two coaxial co- or counter-rotating cylinders -- 
is one of the paradigmatic systems in physics of fluids.
The (dimensionless) control parameters are the Reynolds numbers of the inner and   outer cylinder, the ratio
of the cylinder radii, and the aspect ratio. The response of the system is the torque required to 
retain constant angular velocities, which can be connected to the angular velocity transport through the gap.
While  the low Reynolds number regime has been very well explored in the '80s and '90s of the last century,
in the fully turbulent regime major research activity only developed in the last decade. In this paper we
 review this recent  progress in our understanding of fully developed Taylor-Couette turbulence,  from the
 experimental, numerical, and theoretical point of view. We will focus on the parameter dependence  of the
 global torque and on the local flow organisation, including velocity profiles and boundary layers.
 Next, we will discuss transitions between different (turbulent) flow states.
 We will also elaborate on the relevance of this system for  astrophysical disks (Keplerian flows).  
The review ends with a list of challenges for future research on turbulent Taylor-Couette flow.\\
{\bf The published version in ARFM: Annu. Rev. Fluid Mech. 48:53-80 (2016).}
\end{abstract}


\maketitle

\section{Introduction}\label{intro}

\begin{figure}[h]
\begin{center}
\includegraphics[width=1\textwidth]{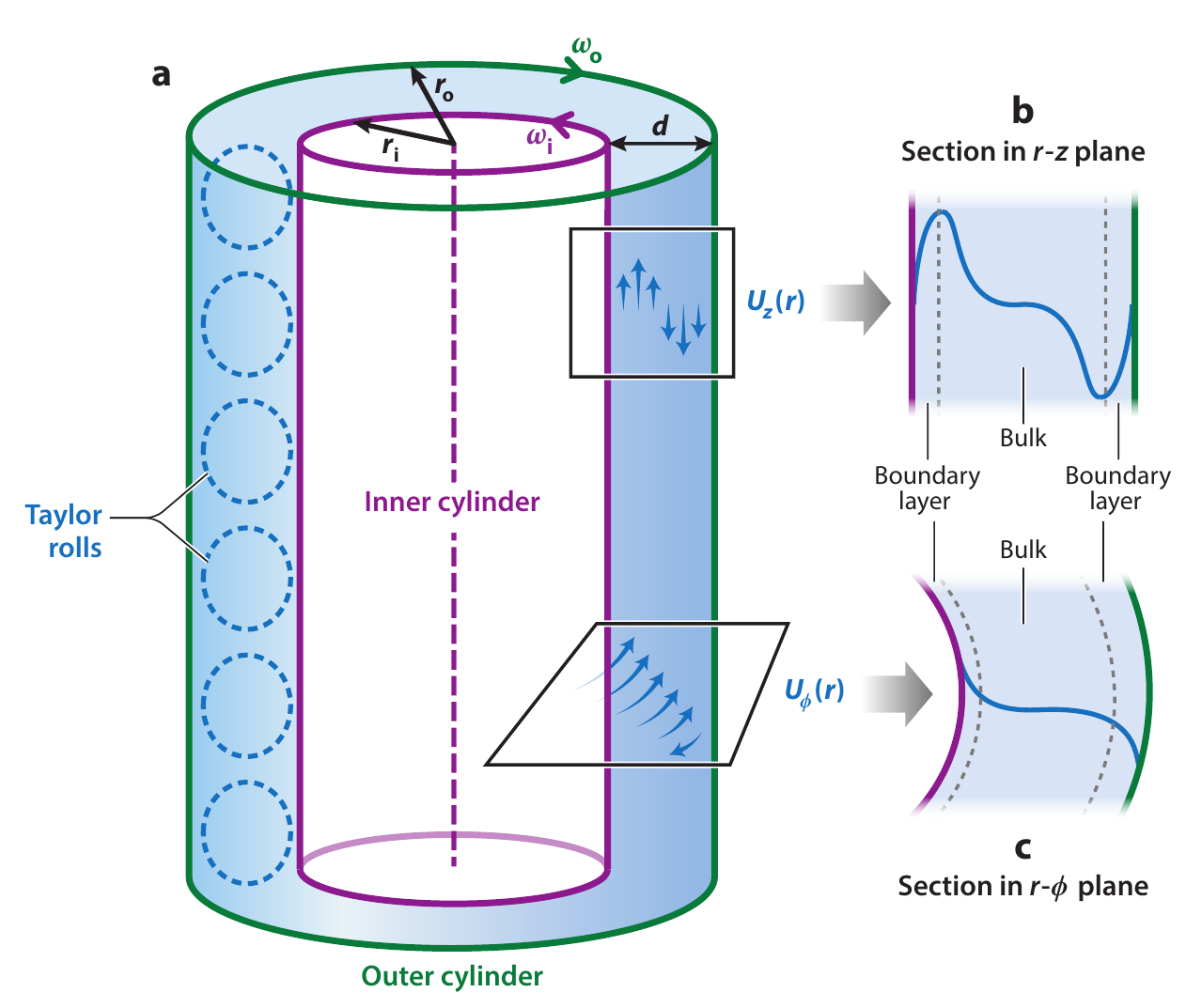}
  \caption{(a) Sketch for Taylor-Couette flow and the notation used in the present work. 
    The inner and outer cylinder radii are $r_i$ and $r_o$, respectively, their respective angular velocities are $\omega_i$ and $\omega_o$.  The gap width is $d=r_o-r_i$. Radial distances from the origin are called $r$. The mean azimuthal velocity field is $U_\phi (r) = r \Omega (r) $, where $\Omega(r)$ is the angular velocity. The mean axial velocity field is $U_z (r)$, which besides on $r$ will also depend on the axial position $z$. The dashed rolls indicate the Taylor roll remnants with axial and radial velocity components, which are considered to be the largest eddies of the turbulent TC flow. A sketch of (b) the axial velocity and (c) the azimuthal velocity profile in the boundary layers (BL) and the bulk.}
\label{fig:TC-sketch}
\end{center}
\end{figure}

Taylor-Couette (TC) flow, the flow between two coaxial, independently rotating cylinders (as sketched in figure~\ref{fig:TC-sketch}), is one of the paradigmatic systems of physics of fluids. It is, next to Rayleigh-B\'enard (RB) flow
(the flow in a box heated from below and cooled from above), the `drosophila' of the field, 
and various new concepts 
in fluid dynamics have been  tested with these systems, be it 
 instabilities 
\citep[e.g.][]{tay23b,col65,cha81,dra81,pri81,bus67,cou96,mar97,hri02,mes02,don91,geb93,tuc14,mar14}, nonlinear dynamics and spatio-temporal chaos 
 \citep[e.g.][]{lor63,ahl74,pfi81,smi82,beh85,mul87,str94},
pattern formation \citep[e.g.][]{cro93,and86,kos93,buc96,bod00},
or turbulence  
\citep[e.g.][]{sig94,kad01,lat92,ahl09,loh10}. 
The reasons why these systems are so popular are from our point of view: 
(i) They  are mathematically well-defined
by the Navier-Stokes equations with their respective
boundary conditions. (ii) For these closed system
exact global balance relations between the respective driving and the 
dissipation can be derived. (iii) They are experimentally accessible with high precision,
thanks to the simple geometries and high symmetries. 
(iv) The boundaries and  resulting boundary layers play a prominent role and 
 thus they are ideal systems to study the interaction between boundary layers and bulk.
(v) There is a close analogy of RB and TC flow with  pipe flow \citep[as e.g. elaborated in][]{eck00,eck07a},
which from a technological point of view may be the most important turbulent flow. So insight into the 
interaction between boundary layers and bulk in TC and RB turbulence will clearly  also shed more light on the
pipe flow problem
\citep[see e.g.][]{zag98,mar10b,hul12}.

\begin{figure}
\begin{center}
\includegraphics[width=1\textwidth]{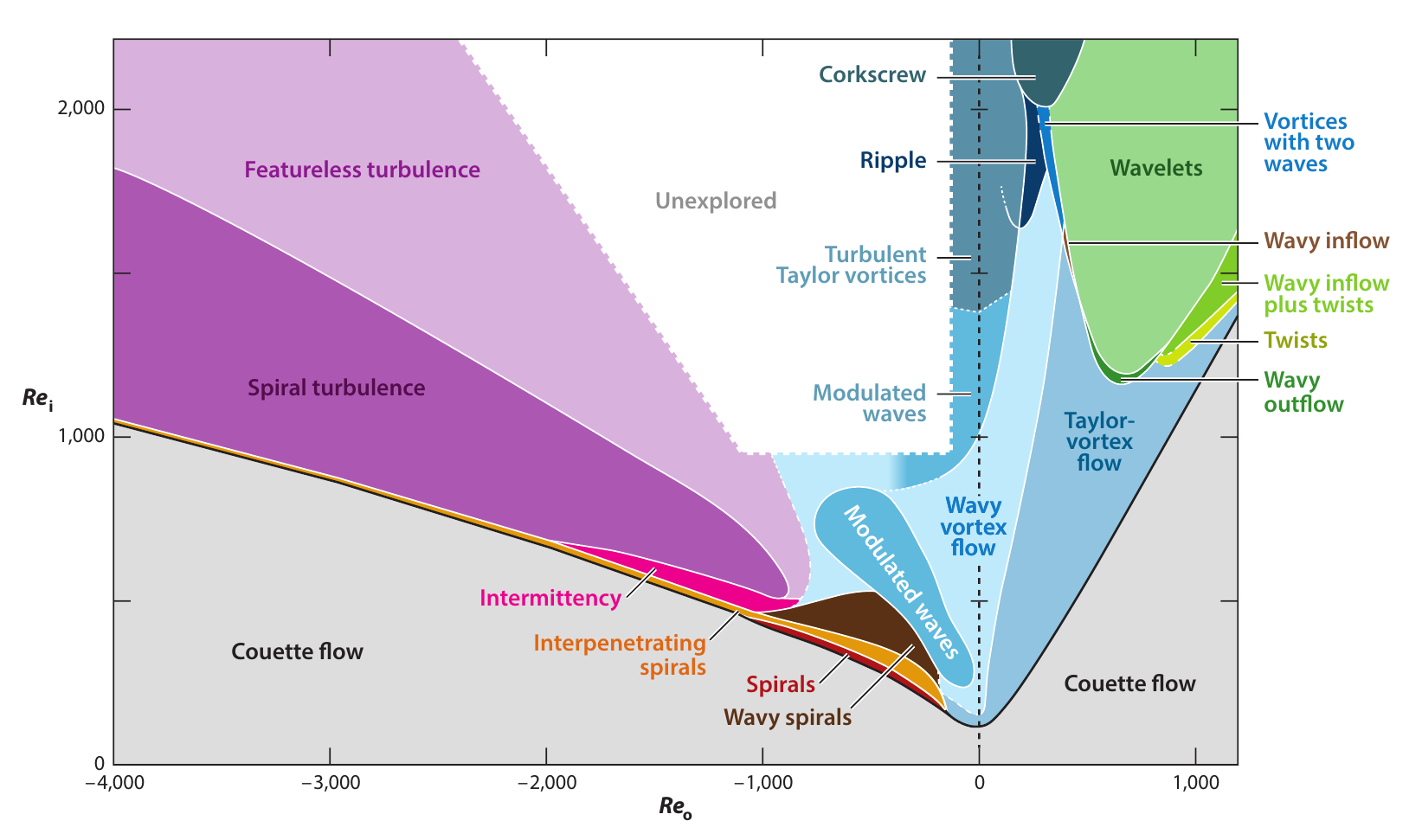}
  \caption{
 Observed rich flow structures in the   ($Re_o$,$Re_i$) phase diagram for TC flow at $\eta = 0.833$.
 Figure taken from \cite{and86}, with permission from the authors.}
\label{fig:andreck-pd}
\end{center}
\end{figure}

In the 1980s much  research was done on TC flow for small Reynolds numbers, at onset 
of the instabilities and
slightly above, in parallel to related work for RB flow. In this regime -- in spite of the low Reynolds numbers up to only
1000 or 2000 -- 
the flow structure is extremely rich, as reflected in the phase diagram of \cite{and86}, which we reproduce in 
figure~\ref{fig:andreck-pd}.
In a nutshell, TC  flow is linearly stable for outer cylinder rotation and fixed or only slowly rotating inner cylinder. 
The onset of instabilities at increasing Reynolds number of the inner cylinder is 
caused by the driving centrifugal force
and can be estimated by force balance arguments as done by \cite{tay23b,ess96}, who generalized  Rayleigh's 
famous stability criterion \citep{ray17}. 
In the unstable regime one can observe Taylor rolls, modulated waves, spirals, and many
other rich spatial  and temporal flow features. 
For more details on this regime we refer to the book by \cite{cho94}, to the recent review by \cite{far14},
 and to the review given in the thesis by \cite{bor14phd}.

While in the 1990s for RB flow the degree of turbulence was  continuously increased  
both experimentally with various setups and numerically with several codes 
\cite[see e.g. the reviews by][]{sig94,ahl09,loh10}, for TC flow there was much less activity, the most visible 
exception being  the Austin-Maryland experiment by \cite{lat92,lat92a,lew99}, who experimentally explored 
fully developed TC turbulence for pure inner cylinder rotation. For {\it independently} rotating cylinders
and at the same time strongly driven  turbulence, 
up to a few years ago the experiments by \cite{wen33} remained the most prominent ones.

\begin{figure}
\begin{center}
\includegraphics[width=1\textwidth]{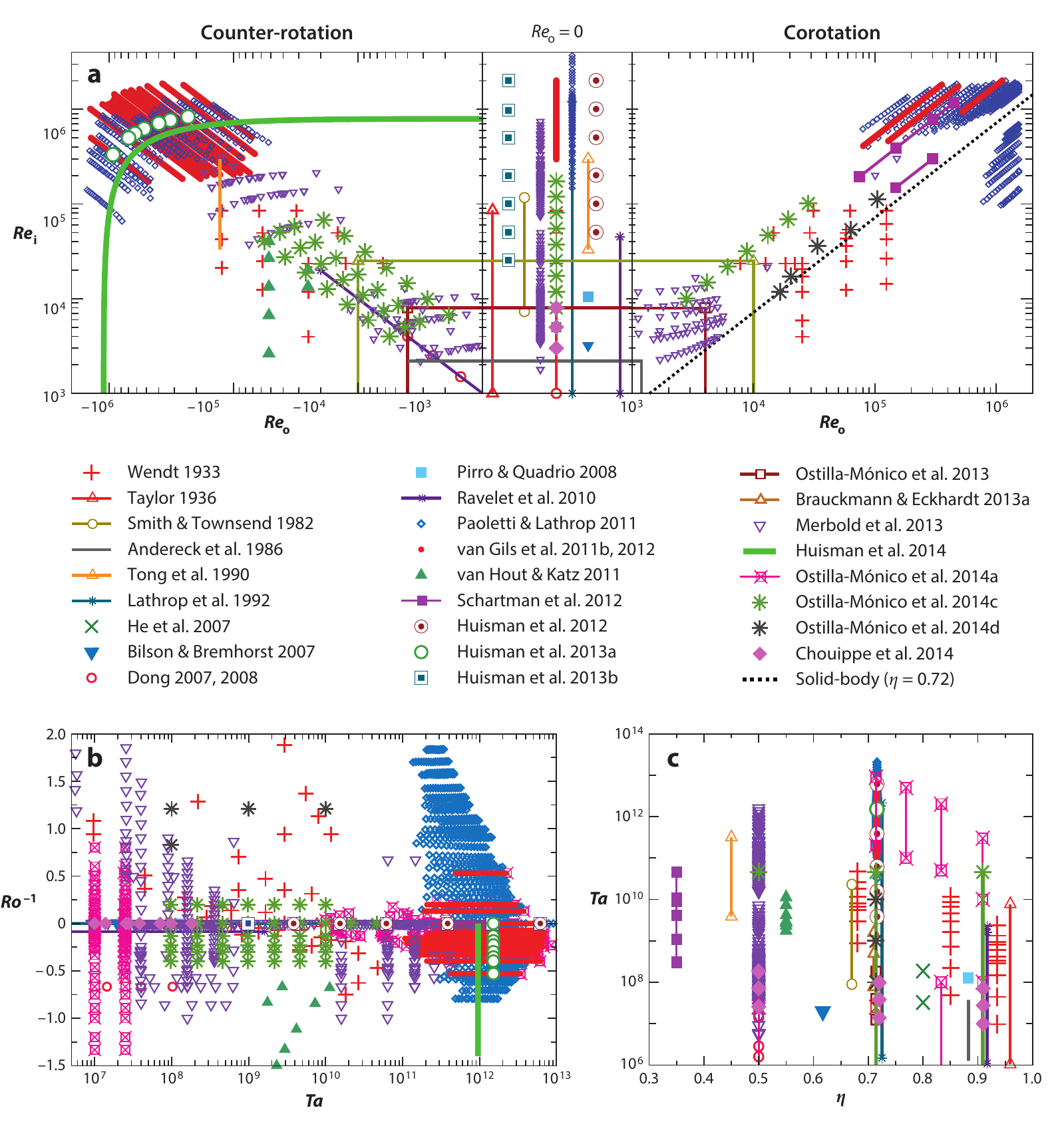}
  \caption{
  (a) Explored ($Re_o$, $Re_i$) parameter space of TC flow with independently rotating 
inner and outer cylinders. Both experimental data \citep{wen33,tay36,smi82,and86,ton90,lat92a,rav10,gil11,gil12,hou11,pao11,schartman12,hui12,mer13,hui13pre,hui13,hui14}
and numerical data \citep{heTC07,bil07,don07,don08,pir08, bra13,ost13,ost14eta,ost14pd,cho14} are shown. The solid-body rotation
line for a radius ratio $\eta=0.71$ is added. Solid lines between markers represent a large density of experiments.
In (b) the same data are shown in the $(Ta, Ro^{-1})$ parameter space and in (c) the explored $(\eta, Ta)$ parameter space is shown.  
 These phase diagrams 
  clearly indicate
   that the field of high Reynolds number TC flow has been progressing rapidly. }
\label{fig:pd}
\end{center}
\end{figure}

This situation 
 has only changed in the last decade, when researchers started to explore  
 the full phase space of TC, i.e.\ also for independently rotating cylinders and in the strongly turbulent regime, 
  well beyond the onset of chaos. 
Many  new experimental setups  have been   built to examine various aspects of 
strongly turbulent TC flow with co- and counter-rotating cylinders 
 \citep{ji06,rav10,bor10,gil11a,gil11,gil12,hou11,pao11,schartman12,hui12,mer13,hui13pre,hui13,hui14} and correspondingly 
 numerical simulations have been 
  pushed to very 
  high Reynolds numbers \citep{heTC07,bil07,don07,don08,pir08, bra13,bra13b,ost13,ost14eta,ost14bl,ost14pd,cho14}. Figure~\ref{fig:pd} shows the currently explored parameter space 
  in different representations. 
  These phase diagrams 
  clearly indicate
   that the field of high Reynolds number TC flow has been progressing rapidly and it is the
  aim of this article to review this progress. 
  
The review is organized as follows: In section \ref{param} we introduce the control parameter space of TC flow
and its global response. 
A general overview of the flow organization in different regimes of the parameter spaces is given in section 3, together 
with the central idea of  how to calculate the global response -- namely a partitioning 
of the flow into  a boundary layer and a bulk part entirely   analogously  to what has been  done in RB flow \citep{gro00,gro01,gro02,gro04}. 
In section \ref{global} we report on the global response of the system. The detailed local flow organization -- profiles and rolls -- is described in section \ref{local},
resulting in  an explanation of when and why the system displays optimal transport properties.
In that section we will also report on the occurrence of multiple turbulent states in TC flow. 
While these first sections focus on the Rayleigh unstable regime, we touch upon the Rayleigh stable regime in section
\ref{kepler}, due to its relevance for astrophysical models. The review ends with a summary and an outlook on open issues
(section \ref{open}).

\section{Control parameters and global response of Taylor-Couette flow}\label{param}

The geometric parameters describing the TC system are 
the inner and the outer cylinder radii $r_i$ and $r_o$, respectively,  the corresponding  gap width  $d= r_o- r_i$,
 and the height
 $L$ of the sample, see figure~\ref{fig:TC-sketch}. 
In dimensionless form these parameters are expressed via the  
 radius ratio $\eta = r_i / r_o$ and  the aspect ratio $\Gamma = L / d$. 
The driving of the system is through rotation of the inner and outer cylinder. In dimensional form this is quantified
by the  angular velocities $\omega_i$ and $\omega_o$, and in dimensionless form by the respective 
Reynolds numbers $Re_i$ and $Re_o$, namely
\be
Re_{i,o} = { r_{i,o} \omega_{i,o} d  \over \nu  } ,
\ee
where $\nu$ is the kinematic viscosity of the fluid in between the cylinders. The convention is that $Re_i$ is always 
positive, whereas $Re_o>0$ stands for a co-rotating  and $Re_o<0$  for a counter-rotating outer cylinder.

Instead of using $Re_i$ and $Re_o$, alternatively the driving of  TC flow can be characterized
by the Taylor number 
\be
Ta = \frac{(1+\eta )^4}{64 \eta^2}  ~ { (r_o - r_i)^2 (r_i+r_o)^2 (\omega_i - \omega_o)^2 \over  \nu^{2}},
\label{eq:Taylor}
\ee
which can be seen as the non-dimensional  differential rotation of the system, and the 
(negative) rotation ratio $a = - \omega_o/\omega_i$, with $a>0 $ for counter-rotation and $a<0$ for co-rotation. 
Alternatively to the rotation ratio one can also use the inverse Rossby number 
\begin{equation}
Ro^{-1} =\displaystyle\frac{2\omega_od}{|\omega_i-\omega_o|r_i}= -2\frac{1-\eta}{\eta}\frac{a}{|1+a|}.
\label{eq:Rosdef}
\end{equation}
The advantage of this  representation is that in the coordinate system co-rotating with the outer cylinder, 
$Ro^{-1}$ directly characterises the strength of the driving Coriolis force, as can be seen from the underlying
Navier-Stokes equation formulated in that coordinate system, 
\begin{equation}
 \displaystyle\frac{\partial {\bu}}{\partial {t}} + {\bu}\cdot{\nabla}{\bu} = -{\nabla} {p} +  
\displaystyle\frac{f(\eta)}{Ta^{1/2}} {\nabla}^2{\bu} - Ro^{-1} {\boldsymbol{e}_z}\times{\bu}~,
\label{eq:TC_NS}
\end{equation}
where $f(\eta) = (1+\eta)^3 / (8\eta^2)$. 
The same data of the $(Re_o, Re_i)$ parameter space of figure~\ref{fig:pd}a are shown
 in figure  \ref{fig:pd}b in the $ (Ta, Ro^{-1})$ parameter space. But as we will see,
 the representation of the data in terms of $Ro^{-1}$ also has
 disadvantages, and we will use $Ro^{-1}$ and the (negative) rotation ratio $a$ in parallel. 

The advantage of using the Taylor number rather than the Reynolds numbers is that then the analogy
between TC flow and RB is illuminated, as worked out in detail in \citet*{eck07b}. 
In that paper it was shown in particular 
that the conserved transport quantity in TC flow is the 
 angular velocity flux from the inner to the outer cylinder, 
\begin{equation}
J^\omega = r^3 \left ( \left< u_r \omega \right>_{A,t} - \nu \partial_r \left < \omega \right >_{A,t} \right ). 
\end{equation}
Here, $u_r$($u_\phi$) is the radial (azimuthal) velocity, $\omega = u_\phi/r$ the
angular velocity, and $\left< \dots\right>_{A,t}$ characterizes averaging over time
and the area with constant distance  from the axis. 
 In analogy to the definition of the dimensionless heat flux in RB flow, \citet*{eck07b} defined a ``Nusselt number'' 
  as the ratio of the angular velocity flux $J^\omega$ and its value $J^\omega_{lam} = 2\nu r_i^2 r_o^2
(\omega_i - \omega_o)/ (r_o^2 - r_i^2)$ for the laminar case, i.e. 
\be
Nu_\omega = J^\omega / J^\omega_{lam}. 
\ee
$Nu_\omega$ is the key response parameter of the TC system. It is directly connected to the torque $\tau$ that is necessary to keep the  angular velocities constant.  Its  dimensionless form  is defined as
\be
G = { \tau\over 2\pi \ell \rho \nu^2 } = Nu_\omega {J^\omega_{lam}\over  \nu^2 } =
Nu_\omega G_{lam}. 
\label{g}
\ee
Here $\ell$ is the height of the part of the cylinder on which the torque is measured, $\rho$ the density of the fluid, and $G_{lam}$ is the dimensionless torque for the laminar case. Also \cite{dub02} worked out the analogy between TC and RB flow, but still used
the dimensionless torque $G$ as response parameter. 
Yet another  possibility often used to represent the data is the friction coefficient
$c_f = ((1-\eta)^2/\pi)G/Re_i^2$ \citep{lat92a}.

The second key 
response parameter is the degree of turbulence of the {\it wind} in the gap of the cylinders, which measures the strength of the secondary flows -- 
 the $r$- and $z$-components of the velocity field ($u_r$ and $u_z$). Since the time-averaged wind velocity is generally very small when the secondary flows are not stable over time --
  i.e. $\left < u_r \right>$ $\simeq$ 0 and $\left < u_\phi \right>$ $\simeq$ 0 --
   the standard deviation of the wind velocity is used  to quantify the wind Reynolds number, 
\be
Re_w = \frac{\sigma_{u_w}(r_o-r_i)}{\nu}. 
\ee
$\sigma_{u_w}$ can be chosen to be the standard deviation of either the radial or axial velocity. 

The key issue now is  to understand how the response parameters $Nu_\omega$ and $Re_w$ depend on 
the control parameters 
 $Ta$, $Ro^{-1}$, and $\eta$. 
 
 \begin{figure}[h]
\begin{center}
\includegraphics[width=1\textwidth]{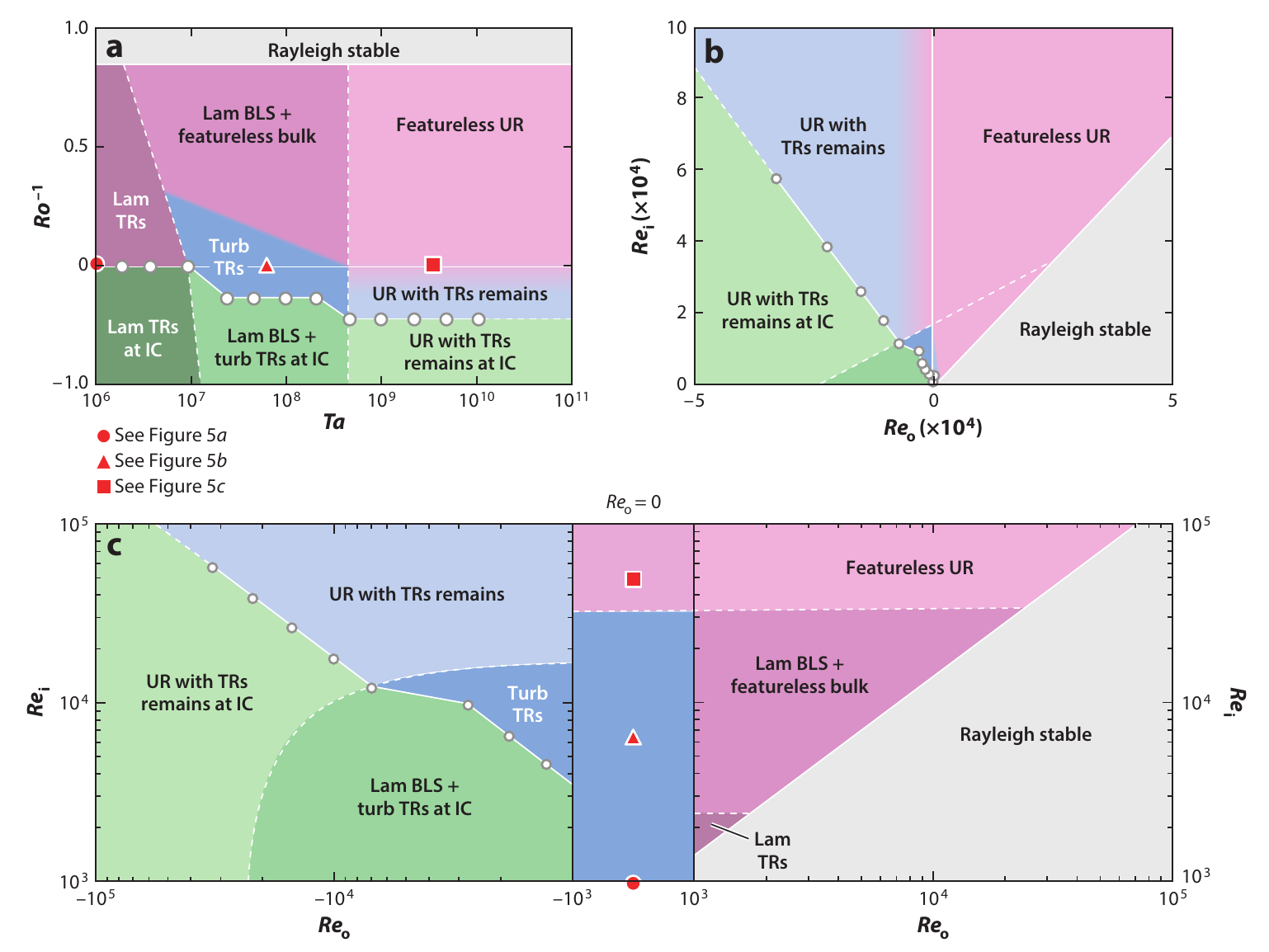}
  \caption{Different regimes in the ($Ta$, $Ro^{-1}$) (top-left) and ($Re_o$, $Re_i$) (top-right and bottom) phase 
  spaces for $\eta =$ 0.714, as obtained from the direct numerical simulations  of \cite{ost14pd}.
   The hollow circles indicate the location of optimal transport, and serve as an indication 
  of the location of the borderline  between the 
  co-rotating or slowly counter-rotating regime (CWCR, blueish and reddish) and 
  the strongly counter-rotating regime (SCR, greenish).
Abbreviations: boundary layer (BL), Taylor rolls (TR), ultimate regime (UR), and inner cylinder (IC). 
Figures taken from \cite{ost14pd}. The red dot, triangle, and square  correspond to the locations of the three cases shown in figure 5.
  }
\label{fig:ostilla-pd}
\end{center}
\end{figure}

 \section{General flow features with increasing driving strength}\label{general} 
 Before we answer this question in section \ref{global}, we will first report the changes in the general flow features
 with increasing degree of turbulence. 
 A good overview is obtained from the direct numerical simulations (DNS) of the Navier-Stokes equation of 
 \cite{ost14pd}, who
 significantly extended the parameter  space of \cite{and86} (figure \ref{fig:andreck-pd}) towards 
  much larger  Reynolds numbers (up to $10^5$), as shown via  phase diagrams  in 
   both the ($Re_o, Re_i$) and the $(Ta, Ro^{-1})$ 
   representation in figure~\ref{fig:ostilla-pd}. 
   Note that the whole phase diagram of figure \ref{fig:andreck-pd} with all its rich structure 
    fits into a dot-like region in the representations of
   figure~\ref{fig:ostilla-pd}b, whereas in the logarithmic representation of figure~\ref{fig:ostilla-pd}a it 
    appears  in the very left part for small 
    $Ta \lesssim 10^6$ 
    as ``lam TRs", ``lam TRs at IC", and of course as    ``Rayleigh stable".

\begin{figure}[h]
\begin{center}
\includegraphics[width=1\textwidth]{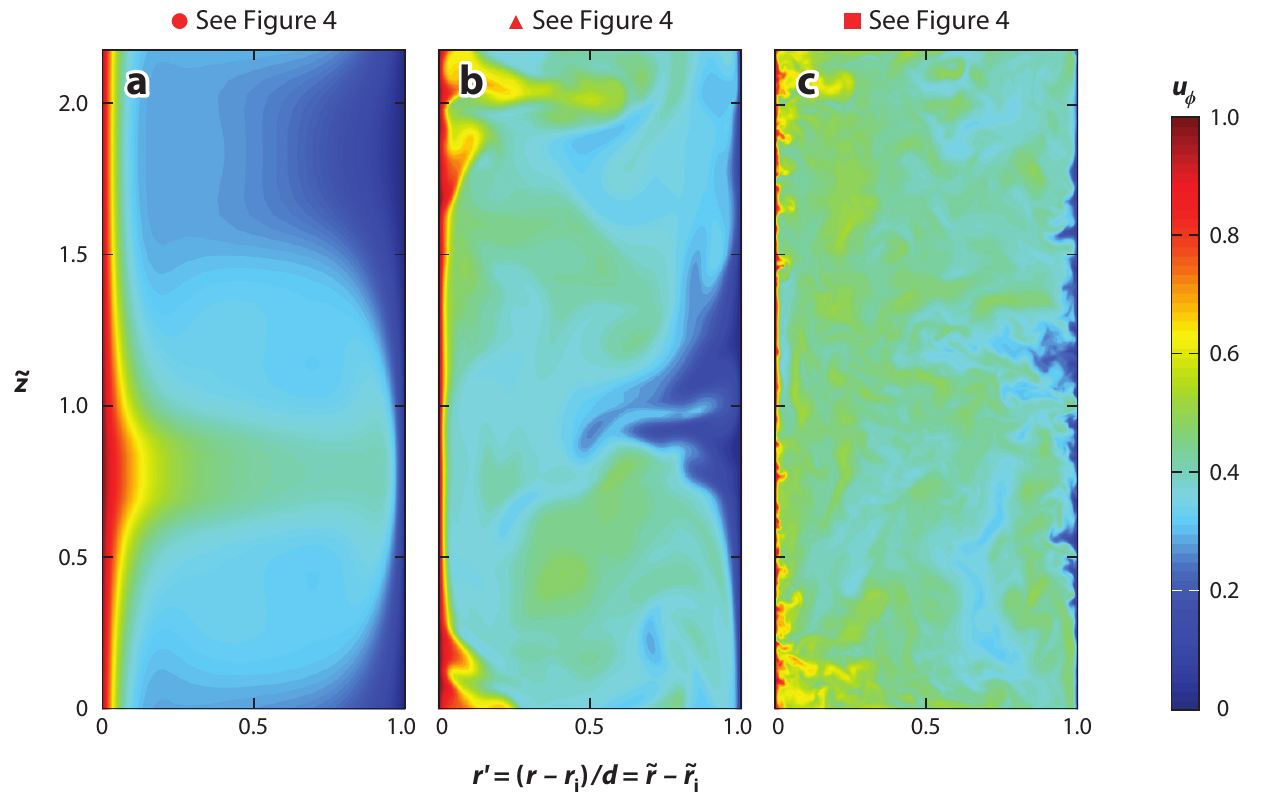}
  \caption{Snapshots of the angular velocity for pure inner cylinder rotation $Ro^{-1} =0$ and $\eta = 0.714$ as obtained from the DNS of \cite{ost14bl}. 
  Periodic boundary conditions in axial direction were used and $\tilde r = r/d$ and $\tilde z = z/d$. 
  (a) $Ta = 7 \times 10^5$ (the red dot in figure~\ref{fig:ostilla-pd}, laminar Taylor rolls), (b)  $Ta = 5 \times 10^7$ (the red triangle in figure~\ref{fig:ostilla-pd}, classical regime with BLs of Prandtl-Blasius type and a turbulent bulk with Taylor rolls),
   (c)  $Ta = 4 \times 10^9$ (the red square in figure~\ref{fig:ostilla-pd}, ultimate regime with turbulent BLs and a featureless turbulent bulk).
    }
\label{fig:snapshots}
\end{center}
\end{figure}

 At  Taylor numbers that are still low ($Ta \sim  10^6$, see figure \ref{fig:snapshots}a), the gap 
 between the cylinders is 
filled with coherent structures (`Taylor rolls', also called 'Taylor vortices')\cite[see e.g.\ the reviews by][]{pri81,tag94,far14}, whose length scale decreases with increasing Ta. 
Around Ta $\approx 3 \times 10^6$ (in the Rayleigh unstable regime) the 
coherence
 length of the structures becomes smaller than the characteristic integral length scale (the gap width $d$) and 
  turbulence starts to develop in the bulk at 
   length scales between the inner (connected to the Kolmogorov scale) and the outer length scale $d$.
    At the same time boundary layers (BLs) start to develop at
     the inner and outer cylinder, with 
         statistical properties 
      that differ  to the bulk flow. 
      The boundary layers 
      are still of laminar type, which -- in spite of time dependence --
      can be described by the Prandtl-Blasius theory. 
      In analogy to RB flow we call this regime -- with  a turbulent bulk and boundary layers of Prandtl-Blasius type --  
       the {\it classical regime} of TC turbulence. A snapshot of the angular velocity profile in this regime is shown in figure \ref{fig:snapshots}b. 
      In the turbulent bulk, depending on $Ro^{-1}$, either the Taylor rolls survive,
      or they partly survive close to the inner cylinder, or else the bulk is featureless, see figure~\ref{fig:ostilla-pd}a and the 
      detailed discussion in section \ref{local}.

  However, 
 around $Ta^*  \simeq 3 \times  10^8$ (for $\eta = 0.71$) the situation changes drastically, 
 as then the boundary layers are sheared strongly enough to  
  undergo a shear instability and become turbulent, i.e., of Prandtl-von K\'arm\'an type. A snapshot of  this regime is shown in figure \ref{fig:snapshots}c. 
 This regime is called the  {\it ultimate regime} of TC turbulence, again in analogy to
 the ultimate regime in RB flow \citep{kra62,ahl09,gro11}. 
 The  typical characteristic of the turbulent boundary layer
 is   the approximately 
  logarithmic velocity profile, as will be  discussed in section 
  \ref{local}. Note that at this transition,  the bulk flow does not change much: Depending on 
  $Ro^{-1}$, it
 is  still  either featureless (for positive $Ro^{-1}$ meaning co-rotation)
  or features Taylor rolls -- either over the full bulk area  (the gap width 
  minus the boundary layers) or, due to the
   stabilisation by strong counter-rotation of the outer cylinder,  only close to the inner cylinder; 
  again see figure~\ref{fig:ostilla-pd}. 
  In addition, the statistics of the     
 turbulent fluctuations in the bulk change only quantitatively 
  beyond the transition towards the ultimate regime, with the inertial regime increasing in size.

 The separation of the flow domain into bulk and BLs is analogous to the situation in RB flow. Therefore 
 the unifying scaling theory developed by \cite{gro00,gro01,gro02,gro04} for RB flow is also applicable here. 
 This was done by  \citet*{eck07b}. Here we only note that this also holds for the transition to the 
 ultimate regime and for the ultimate regime itself, 
 both for the scaling of $Nu_\omega$ and for the profiles \citep{gro11,gro12,gro14}, as we will elaborate
 in the next two sections. 
 
 There is,  however,  a quantitative difference: 
 In RB flow the shear instability of the kinetic boundary layer is only indirectly 
 induced by the thermal driving; namely, the driving first induces a large scale wind, which then in turn builds up the shear near the boundaries. 
The $Ra$ range in the classical state is very wide ($Ra$ from 10$^7$ to 10$^{14}$), and the system only enters into the ultimate state at extremely high $Ra$ number ($Ra^* \simeq 10^{14}$ for Prandtl number $\sim$ 1) that results in many challenges for experiments and numerics \citep{ahl09}. Though numerous efforts have been put forward to reach this critical Rayleigh number $Ra^*$ in order to study the ultimate RB turbulence, only very few experiments reach this value, e.g.\ those of  \cite{he12,he12a,roc10,ahl14}.

In contrast, in  TC flow  the driving is directly  through 
 the rotating cylinders, giving rise to  very large shear directly. 
Since the shear driving in TC flow is much more efficient than the indirect driving by heating in RB flow, 
the TC flow enters the ultimate regime at a much lower $Ta^* \simeq 3 \times 10^8$ (for $\eta = 0.71$). Therefore
 it is an ideal system to study  turbulent flow in the ultimate regime at a wide $Ta$ regime ($10^8$ to $10^{13}$, the latter of which is
 what presently can be achieved in the T$^3$C setup). 
In the next  section, we will focus on the 
dependence of $Nu_\omega$ on the control parameters ($Ta$, $a$, and $\eta$) in both the classical 
(up to $Ta^*  \simeq 3 \times 10^8$)
and the 
ultimate regime (beyond $Ta \gg 3 \times 10^8$).

\begin{figure}
\begin{center}
\includegraphics[width=1\textwidth]{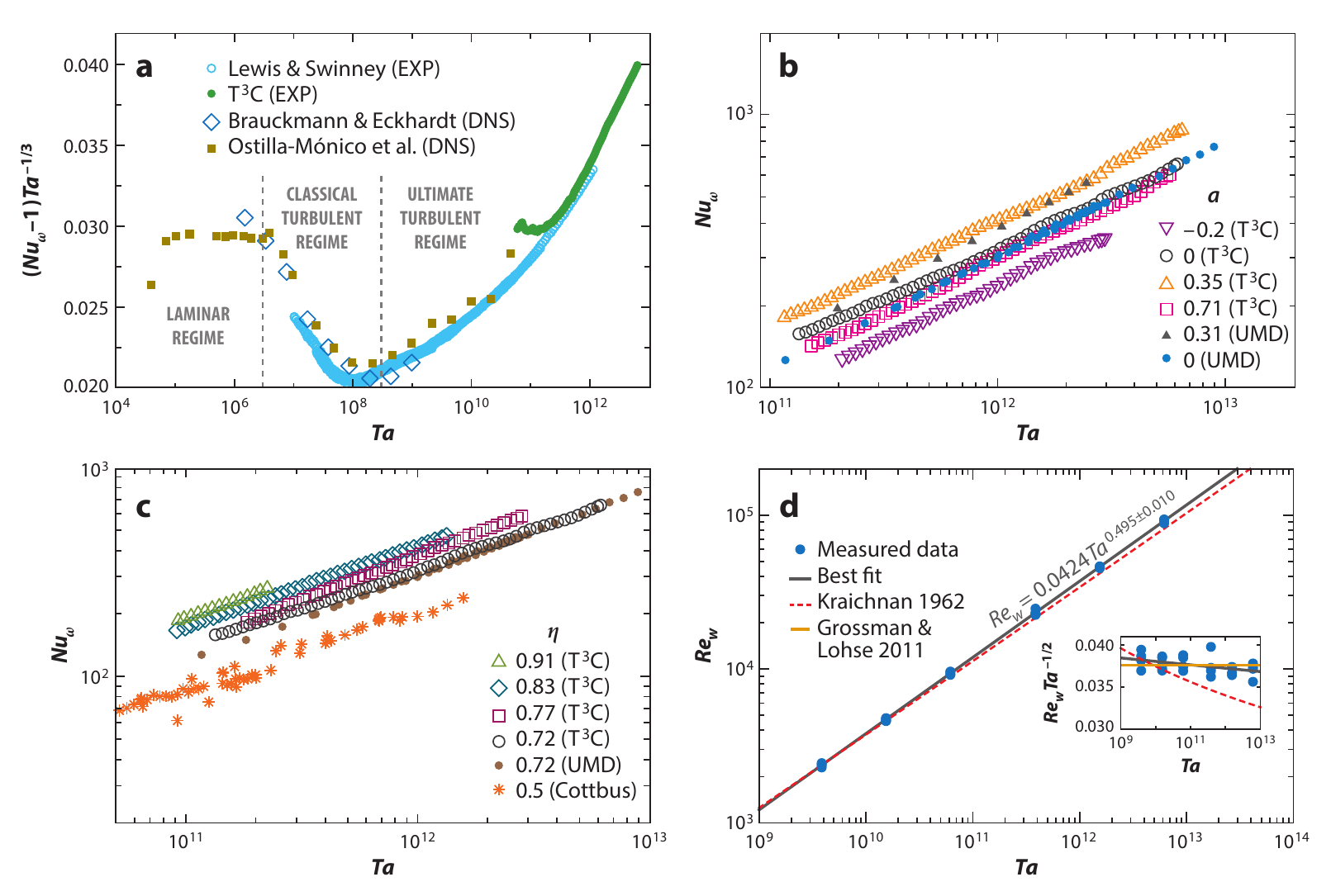}
  \caption{
  (a)  Compensated $Nu_\omega$ vs.\  $Ta$ for $\eta =$ 0.71$-$0.72. 
  The data are from experiments ($\eta = 0.72$) 
  by \cite{lew99} and \cite{gil12} (T$^3$C) and numerical simulations ($\eta = 0.71$)
   by \cite{bra13} and \cite{ost14bl}. Figure taken from \cite{ost14bl}. 
(b) $Nu_\omega$ vs.\  $Ta$ at various rotation ratios $a$, measured in two different facilities for $\eta \simeq $0.72: 
 T$^3$C - data measured in the Twente Turbulent Taylor-Couette facility \citep{gil11,gil12}, 
 UMD -  data measured in the Maryland TC facility \citep{pao11}.  
(c) The measured data on $Nu_\omega$ vs.\  $Ta$ for only inner cylinder rotation ($a$ = 0) at various radius ratios $\eta$: 
T$^3$C - data measured in the Twente Turbulent Taylor-Couette facility \citep{gil11,gil12}, 
UMD -  data measured in the  Maryland TC facility \citep{pao11}, 
Cottbus - data measured in the
 Cottbus TC facility \citep{mer13}.
  (d) The wind Reynolds number $Re_w$ vs. $Ta$. The data measured at midheight in the T$^3$C facility are shown as separate blue dots, showing the quality of the reproducibility and the statistical stationarity of the measurements. The straight line is the best fit $Re_w = 0.0424 Ta^{0.495\pm0.010}$ and the red dashed line is the \cite{kra62} prediction. The inset shows the compensated plot $Re_w/Ta^{1/2}$ vs.\  $Ta$. The horizontal green line is the prediction by \cite{gro11}. Figure taken from \cite{hui12}. 
}
\label{fig:NuRevsTa}
\end{center}
\end{figure}     

\section{Global response of the TC system: $Nu_w$ and $Re_w$}\label{global}

The transitions between the different regimes in the phase space are reflected in the global responses of the system. 
In this section we will discuss these dependences one for   one. 

\subsection{$Nu_\omega$ and $Re_w$ vs.\ $Ta$}
The most relevant dependence may be that of $Nu_\omega$ on $Ta$. In a log-log plot of these 
quantities it is hard to recognize the transitions. They become much more visible in so-called {\it compensated} plots such as shown
in 
figure~\ref{fig:NuRevsTa}a, in which 
 $(Nu_\omega - 1)/Ta^{1/3}$ is plotted vs.\ $Ta$,  for pure inner cylinder rotation $a = 0$,  for
  radius ratios $\eta$  between 0.71 and 0.72, and for 
 10$^4 < Ta <10^{13}$. 
 By plotting the data in this compensated way (i) the laminar case $Nu_\omega = 1$ is subtracted and (ii) 
 one focuses on the differences between
  the scaling exponents of  the classical regime (where for $a=0$ it is $\le 1/3$) and the ultimate
 regime (where in general  it is larger than $1/3$, but $\le 1/2$, which is the theoretical upper bound \citep{how72,doe94}).
 As seen from figure~\ref{fig:NuRevsTa}a, the experimental and numerical data originating from  two different  experimental setups 
  and two different DNS codes agree very well: 
 After the onset of Taylor vortices at $Ta \simeq 10^4$ and up to $Ta \simeq 3 \times 10^6$, one observes the  scaling law $Nu_\omega - 1 \sim Ta^{1/3}$
 for 
 the pure 
  laminar regime. 
  Then the flow becomes 
  time-dependent 
  and for $ 3 \times 10^6 \lesssim  Ta \lesssim 3 \times 10^8$  
 the  effective  scaling  $Nu_\omega \sim Ta^\gamma$ with $\gamma < 1/3$ of  the classical turbulent state  is found, in correspondence with the classical effective $Nu \sim Ra^\gamma$ scaling in RB turbulence \citep{ahl09}.
 In this regime turbulence starts to develop in the bulk of the gap, whereas the BLs  still remain of laminar 
 type. At some transitional  Taylor number $Ta^*$, which depends on $\eta$ and which for 
 $\eta = 0.71$ is   $Ta^* \simeq 3 \times 10^8$,  the BLs also become turbulent and the flow undergoes a transition 
  to the ultimate regime with an  effective exponent $\gamma > 1/3 $. 
   In fact, in the $Ta$ range of $10^{11}$ to $10^{13}$,
the effective exponent $\gamma$ is about 0.39 $\pm$ 0.02,   which in this regime 
 is consistent with the theoretical prediction of $Nu_\omega \sim Ta^{1/2} \times$ log-corrections by \cite{kra62} and by \citet*{gro11}. 
 Such log-correction  predictions are typical for logarithmic BLs. 
 Here they were originally 
derived for RB flow, but 
 can straightforwardly be  translated to TC flow based on the analogies between the two systems.

How does the scaling change for co- or counter-rotation, i.e., $a\ne 0$, (but fixed $\eta = $ 0.72)? 
Figure \ref{fig:NuRevsTa}b shows $Nu_\omega(Ta)$  measured in two different TC facilities for various $a$.   
These
 measurements suggest that the scaling exponent $\gamma$ of $Nu_\omega \sim Ta^{\gamma}$ has a weak dependence on $a$ in the Rayleigh unstable regime. 
 And what is the dependence on the radius ratio $\eta$? 
Figure~\ref{fig:NuRevsTa}c displays $Nu_\omega$ vs.\  $Ta$ at $a = $0 for various $\eta$ 
for  three different TC facilities. Again the scaling exponent $\gamma$ seems to vary only  weakly. 
In summary, for all examined $a$ and $\eta$ values, the scaling exponent $\gamma$ is found to be about 0.39 $\pm$ 0.03  for
 $10^{11} \le Ta  \le 10^{13}$.

What about the corresponding scaling of the wind Reynolds number? For the ultimate regime (originally in RB flow) 
  \cite{kra62}  had derived  log-corrections as well, namely $Re_w \sim Ta^{1/2}\left(\log Ta\right)^{-1/2}$. 
In contrast, \citet*{gro11} derived a different prediction: the logarithmic corrections due to the viscous (wind) boundary layer and the thermal (azimuthal velocity in TC) boundary layer remarkably cancel out, resulting in the scaling $Re_w \sim Ta^{1/2}$, without  any logarithmic corrections. 
Using high-speed Particle Image Velocimetry (PIV), \citet{hui12} directly measured the 
radial velocity fluctuations and extracted the wind Reynolds number as $Re_w = {\sigma_{u_r}(r_o-r_i)}/{\nu} $, where $\sigma_{u_r}$  is the standard deviation of the radial velocity. The measurements 
 shown in figure \ref{fig:NuRevsTa}d reveal a clear scaling of the wind Reynolds number with the Taylor number, $Re_w \sim Ta^{0.495\pm0.010}$, which is consistent with the prediction % $Re_w \sim Ta^{1/2}$  
by \citet*{gro11}.

 \begin{figure}
\begin{center}
\includegraphics[width=1\textwidth]{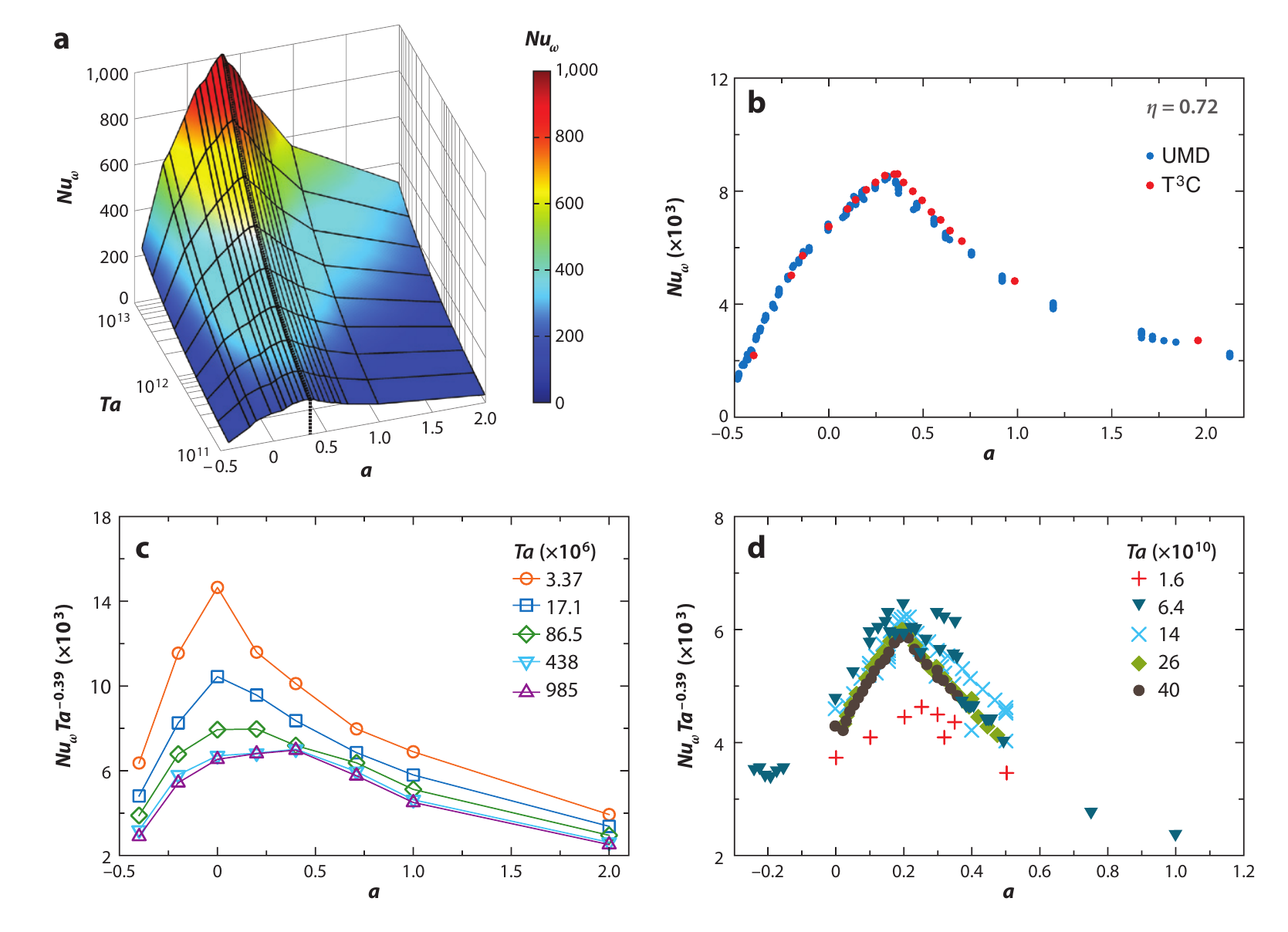}
  \caption{
(a) Three-dimensional (interpolated and extrapolated) overview $Nu_\omega(Ta, a)$ of the experimental results by \cite{gil12}. The color and the height correspond to the $Nu_\omega$ value. Figure taken from \cite{gil12}. (b) The compensated $Nu_\omega/Ta^{0.39}$ vs.\ $a$ for the data measured in two different experimental facilities, i.e. T$^3$C 
 \citep{gil11,gil12} and  Maryland  \citep{pao11}. The two data sets show excellent agreement.
  (c) Numerical data: $Nu_\omega/Ta^{0.39}$ versus rotation ratio $a$ for varying Ta numbers at $\eta =$ 0.5. Figure taken from \cite{bra13}
with permission from the authors. (d) Experimental data: $Nu_\omega/Ta^{0.39}$ versus rotation ratio $a$ for varying Ta numbers at $\eta =$ 0.5. Figure taken from \cite{mer13} with permission from the authors. 
  }
\label{fig:Nuvsa}
\end{center}
\end{figure}

 \subsection{$Nu_\omega$ vs.\ $a$ or vs.\ $Ro^{-1}$}
  From figure \ref{fig:NuRevsTa}b we have seen  that the rotation ratio 
  $a$ hardly affects  the {\it scaling}  of $Nu_\omega$ vs.\  $Ta$. However, this is very different for 
  the {\it absolute value} of $Nu_\omega$, which does depend on $a$.  
Figure~\ref{fig:Nuvsa}a gives  a three-dimensional representation of the measured full dependence
$Nu_\omega (Ta,  a)$, clearly revealing  a non-monotonic dependence of $Nu_\omega$ on $a$ with 
 a pronounced maximum at $a_{opt}$ = 0.33 $\pm 0.03$ (for the radius ratio $\eta = 0.72$ of that specific experiment and for
 large enough $Ta$), 
very different from $a=0$. 
This pronounced maximum of $Nu_\omega$ -- also clearly seen in the projection figure \ref{fig:Nuvsa}b -- 
 reflects 
 the optimal angular velocity transport from the inner to the outer cylinder at this angular velocity ratio. 
 The positions of the maxima in the $(Ta, Ro^{-1}$) and $(Re_o, Re_i)$ parameter spaces is  marked as a connected line of 
 white circles, see figure \ref{fig:ostilla-pd}. For the given radius ratio $a_{opt} = 0.33 \pm 0.03$ corresponds to 
 $Ro^{-1}_{opt} = -0.20 \pm 0.02 $. 
 Intuitively, one might have expected that $Nu_\omega (Ta, Ro^{-1})$ has its maximum at $Ro^{-1}$ = 0, i.e. $\omega_o$ = 0 (pure inner cylinder rotation), since outer cylinder rotation stabilizes an increasing part of the flow volume for increasing counter-rotation rate. 
 This  indeed is the case for small $Ta \lesssim 10^7$, as seen from fig.\ \ref{fig:ostilla-pd} and fig.\ 
 \ref{fig:Nuvsa}c, but for larger $Ta$ 
  weak counter-rotation ($0 < Ro^{-1} < Ro^{-1}_{opt}$) can further enhance the angular velocity transport through 
  intermittent turbulent bursts from the BLs \citep{gil12}, 
  though the flow is predominantly Rayleigh-stable near the outer cylinder. 
Following \cite{ost14pd}, we  call the regime above the line of the maxima $Ro^{-1}_{opt}$ in the parameter spaces of figure  \ref{fig:ostilla-pd}
 the co-rotating or weakly counter-rotating regime (CWCR, shaded blueish and reddish in that figure).
 A reduction of the angular velocity transport only sets in below this line,  in what we call the
  strongly counter-rotating regime (SCR, greenish).
 In general, the line of the maxima in parameter space reflects the  competition of 
  destabilization  by the inner cylinder rotation and stabilization  by the outer cylinder rotation. 
 This competition depends on the radius ratio $\eta$, as seen from figures \ref{fig:Nuvsa}c,d, in which the angular velocity
 transport as measured by  
 \cite{mer13} for $\eta = 0.5$ is  shown, displaying a maximum around  $a_{opt} = 0.20$ to $0.25$
  for large enough $Ta$, corresponding to $Ro_{opt}^{-1} =$ $-$0.33 to $-$0.4.

Note that in figures~\ref{fig:Nuvsa}b-d  we have compensated $Nu_\omega$ with $Ta^{0.39}$.
This is the mean effective scaling of $Nu_\omega$ in the  
 explored large $Ta$ number regime ($10^9$ to $10^{13}$), in which $Nu_\omega (Ta, Ro^{-1}, \eta)$ 
 factorizes into  $Nu_\omega (Ta, Ro^{-1}, \eta) \sim Ta^{0.39} f(Ro^{-1}, \eta )$ or  
  $Nu_\omega (Ta, a , \eta) \sim Ta^{0.39} f(a, \eta )$. In section \ref{local} we will explain the dependence 
  of the scaling function $f$ on $Ro^{-1}$ and $\eta$.

\begin{figure}
\begin{center}
\includegraphics[width=1\textwidth]{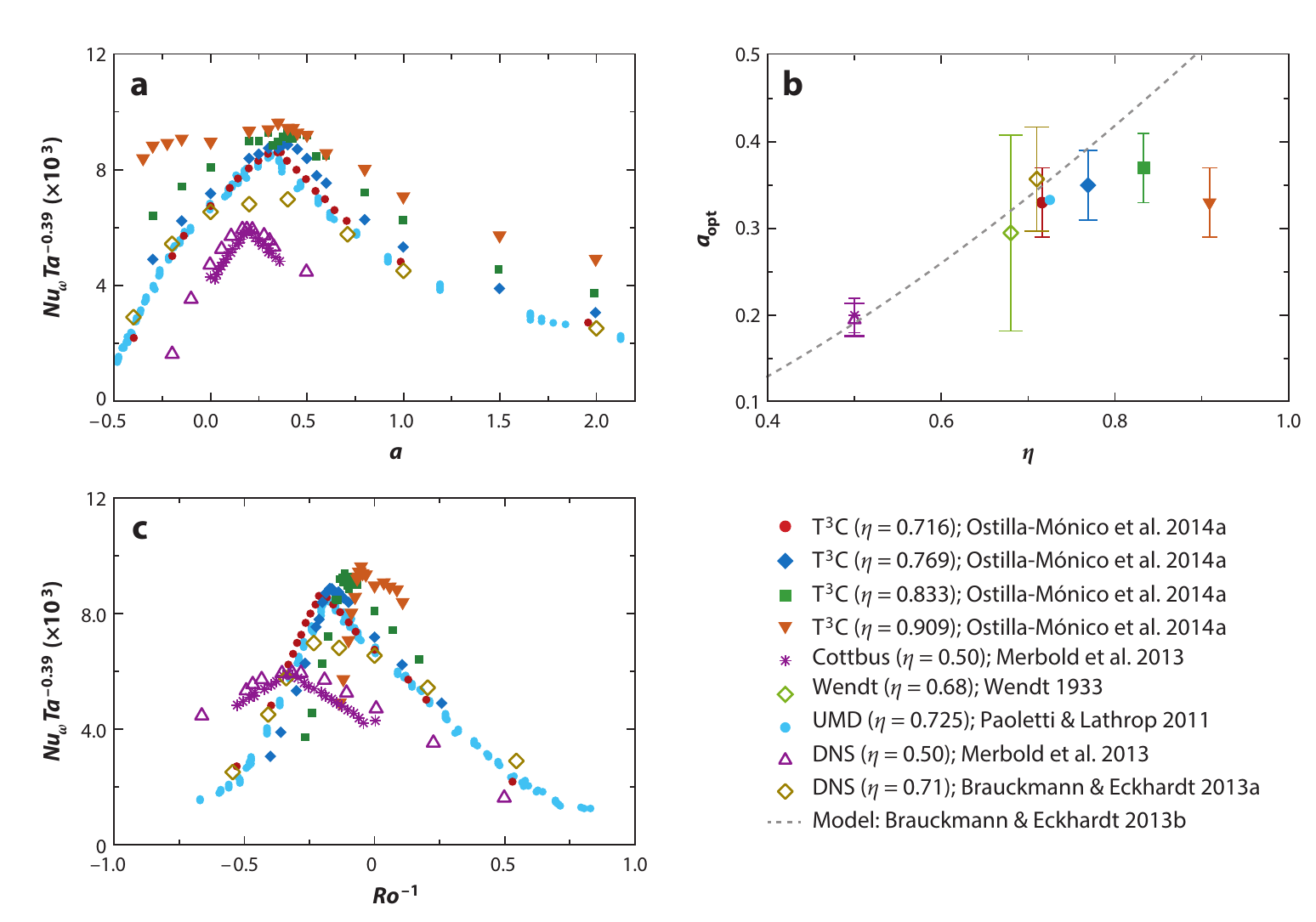}
  \caption{
  Radius ratio $\eta$-dependence of $Nu_\omega$: 
  Plots of $Nu_\omega/Ta^{0.39}$ versus (a) a and versus (c) $Ro^{-1}$ from the collection of 
   state-of-the-art data at various $\eta$. (b) $a_{opt}$ obtained from  figure (a) versus $\eta$. Symbols represent experimental and numerical results; 
  the model prediction of \cite{bra13b} is  plotted as dashed line. Figure taken from \cite{ost14eta}.
      }
\label{fig:Nuvseta}
\end{center}
\end{figure}

\subsection{$Nu_\omega$ vs.\ $\eta$}
We finally report  the radius ratio 
dependence of $Nu_\omega$ in more detail. 
Figure~\ref{fig:Nuvseta}a shows $Nu_\omega/Ta^{0.39}$ as function of $a$ for
 varying values of $\eta$ from 0.5 to 0.909. 
 $Nu_\omega$
  increases with increasing $\eta$ in the present parameter regime, indicating
  that  the angular velocity transport is more efficient for a larger radius ratio $\eta$. As seen  in figure~\ref{fig:Nuvseta}a, the shapes of the
  curves  $Nu_\omega/Ta^{0.39}$ vs.\ $a$  depend strongly 
   on $\eta$: the peak is very narrow 
    and pronounced at low radius ratios, and the location of $a_{opt}$ for the optimal transport can be easily identified.
  In contrast, for larger radius ratios
  the peak of the curve becomes less steep.  This suggests that the angular velocity ratio 
  of optimal transport  $a_{opt}$ becomes less 
  sharply  defined  as the radius ratio $\eta$ approaches  1. As can be seen  from 
   figure~\ref{fig:Nuvseta}a, it is very difficult to precisely identify the location of $a_{opt}$ for $\eta = 0.909$, since the curve is almost flat for the data of $\eta = 0.909$ at $a < 0.5$. 

From the curves in figure~\ref{fig:Nuvseta}a, we can identify  $a_{opt}$ as  a function of $\eta$, 
 which is 
plotted  in figure~\ref{fig:Nuvseta}b. These data assemble 
 the state-of-the-art results on the optimal transport in high-Reynolds number TC flows. Overall, $a_{opt}$ increases with increasing 
$\eta$ for $\eta \lesssim 0.8$, and it seems to saturate at higher $\eta$.

The same data $Nu_\omega (a)$ in figure~\ref{fig:Nuvseta}a are shown  in the   $Nu_\omega(Ro^{-1})$ representation in
 figure~\ref{fig:Nuvseta}c, seemingly exhibiting a different trend: the steepness of the peak does not depend 
 much on the radius ratio $\eta$. This difference in the trends in figure~\ref{fig:Nuvseta}a and figure~\ref{fig:Nuvseta}c is due to the transformation between the control parameters 
  $a$ and $Ro^{-1}$ (eq.~\ref{eq:Rosdef}). As we know, $Ro^{-1}$ takes its justification as  control parameter 
 directly from the Navier-Stokes equations, in which it
  characterises the strength of the Coriolis force. However, it is hard to compare the data of $Nu_\omega ( Ro^{-1}) $ for different $\eta$, because $Ro$ is defined as the ratio of shear rate
  ($\left( \omega_i - \omega_o \right)/d$) and solid-body rotation ($\omega_o/r_i$), which involve two different length scales, i.e. $d$ and $r_i$.

 We finally also note that the transitional  Taylor number Ta$^*$ to the ultimate state is larger at smaller radius ratios,  e.g.,  Ta$^*\simeq 10^{10}$ for $\eta = 0.5$ \citep{ost14pd}.

\begin{figure}
\begin{center}
\includegraphics[width=1\textwidth]{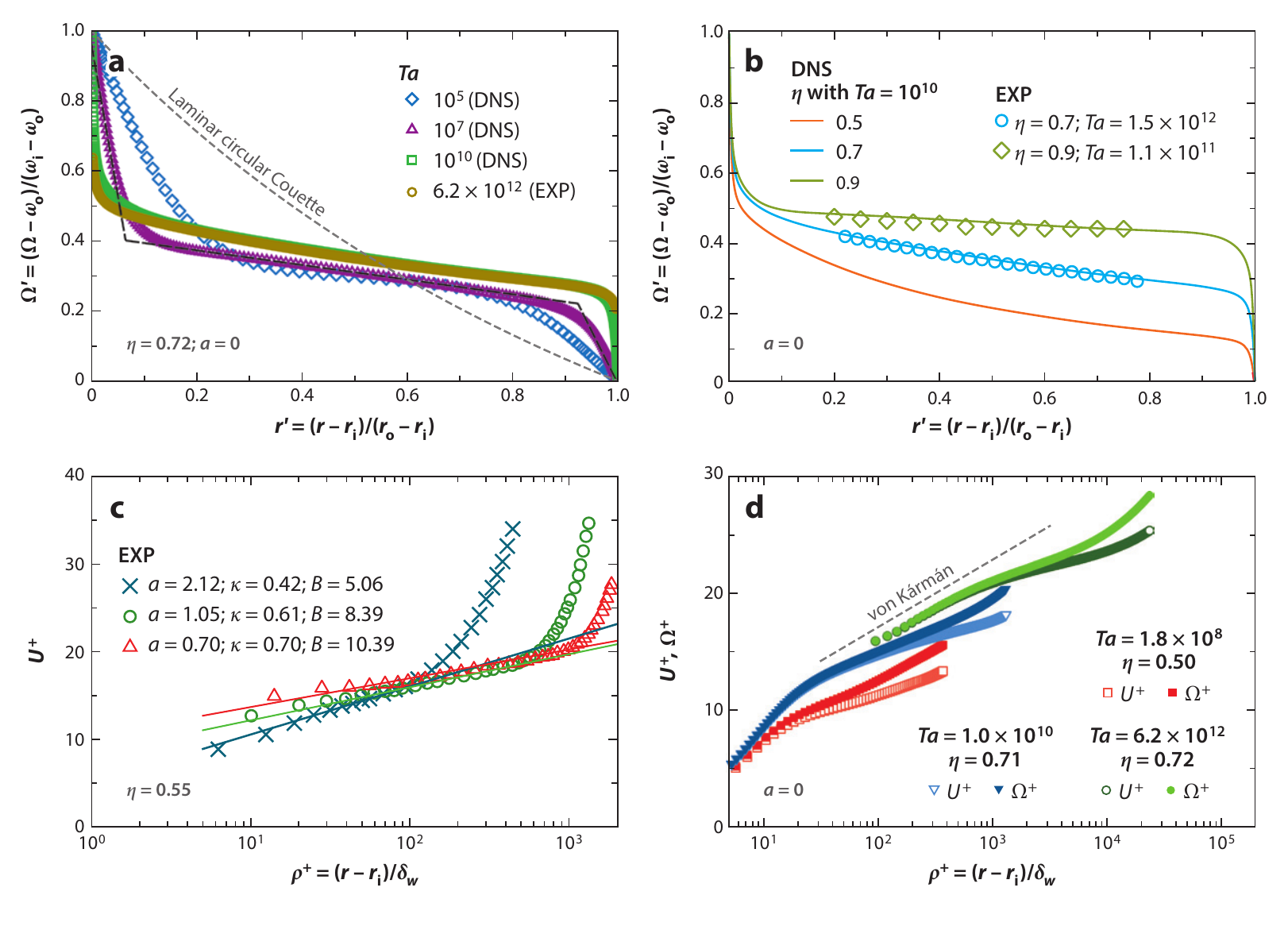}
  \caption{
  (a) Angular velocity profiles for varying Taylor numbers across the TC gap for only inner cylinder rotation at $\eta = $ 0.71 (DNS) and 0.72 (experiments). 
The black line represents the exact laminar circular-Couette (nonvortical) solution of the Navier-Stokes equations. Figure adapted from \cite{hui13}. 
(b) Angular velocity profiles for varying radius ratio across the TC gap for only inner cylinder rotation. The lines are the results from numerical simulations  and the symbols represent  the experimental data. Figure taken from \cite{ost14eta}. 
(c) The inner cylinder azimuthal velocity profile in 
 wall units vs.\ the radial distance from the wall (again in  wall units)
  at varying rotation ratios. Here $\eta =  0.55$ and the
  Taylor numbers are $4.9 \times 10^9$ ($a=$2.12), $7.3 \times 10^9$ ($a=$1.05), and $1.1 \times 10^{10} ($a=$0.70)$, respectively. 
Figure taken from \cite{hou11} with permission from the authors. 
(d) A log-linear plot of the inner cylinder azimuthal velocity profiles ($U^+$) and angular velocity profiles ($\Omega^+$) near the inner cylinder in so-called wall units. Experimental data at $Ta = 6.2 \times 10^{12}$ are taken from \cite{hui13}, 
the numerical data at  $Ta = 1.0 \times 10^{10}$  from \cite{ost14bl}, and the numerical data at  $Ta = 1.8 \times 10^{8}$ 
from \cite{cho14} with permission from the authors. The radial distance from the inner cylinder wall is 
 normalized by the wall unit, i.e. $\rho^+ = (r-r_i)/\delta_{w,i}$, where $\delta_{w,i}$
 is  the viscous length scale for the inner cylinder boundary layer.
}
\label{fig:profile}
\end{center}
\end{figure}

\section{Local flow organization: Profiles,  rolls, and optimal transport}\label{local}
After the  general overview of the flow organization in section \ref{general} and after having reported the global flow response 
in section \ref{global}, we will now look into the profiles and into the detailed flow organization. This will also allow us to 
rationalize 
the dependence $f(Ro^{-1}, \eta )$ of the global transport properties reported above.

\subsection{Profiles} 
For Taylor numbers below the threshold $Ta_c \approx 10^4$ for the onset of instabilities, 
the angular velocity profile in TC flow follows the classical non-vortical laminar profile \cite[see e.g.][]{ll87},
\be
\Omega_{\phi, \mathrm{lam}} (r) = A   + B/r^2  \ , \qquad A = \frac{\omega_o - \eta^2
 \omega_i}{1-\eta^2} \ , \qquad B = \frac{(\omega_i-\omega_o)r_i^2}{1 - \eta^2} \ . 
\label{eq:laminar_profile}
\ee
For larger Ta the flow becomes time dependent and then $\Omega (r) = \left< \omega (r) \right>_t$ is the time-averaged angular velocity.
 This profile is shown as  the black line in figure~\ref{fig:profile}a (for 
$\eta = 0.7$),  in the representation 
 $\Omega' = \left(\Omega(r)-\omega_o\right)/\left(\omega_i-\omega_o\right)$ vs.\  $r'=\left(r-r_o\right)/\left(r_i-r_o\right)$.
  With increasing $Ta$ number,  the profiles start to deviate from the non-vortical laminar one, as 
  also shown in  figure~\ref{fig:profile}a 
  (for the case of zero  outer cylinder rotation, $Ro^{-1} = 0$). 
After the onset of Taylor vortices, eventually time dependences 
set in
 and the large-scale coherent structures break up into smaller structures. Finally, the separation between 
  the bulk and the BLs starts to develop. 
  As seen from  figures~\ref{fig:snapshots}a,b and~\ref{fig:profile}a, 
   between  $Ta = 10^5$ and $10^7$ the profiles 
   can be decomposed into a   bulk  and two boundary layers. 
   The boundary layers can be well characterized by the Prandtl-Blasius 
BL  theory for BLs of    laminar type, 
as illustrated by the dashed lines in the figure. With further increase of  Ta, the BLs become increasingly  
  thin and 
  around $Ta^* \simeq 3 \times 10^8$ 
  finally undergo a transition  to turbulence BLs of Prandtl-von K\'arm\'an type, see 
  figure~\ref{fig:snapshots}c. In figure~\ref{fig:profile}a 
  we show the angular velocity 
  profiles  at $Ta = 10^{10}$ and $6.2 \times 10^{12}$, both clearly being of turbulent type with a logarithmic shape,
  as we will see later.

The angular velocity profile also strongly 
depends on the radius ratio $\eta$,  as seen from the experimental and DNS data  in figure~\ref{fig:profile}b. 
The angular velocity gradient in the bulk decreases with increasing $\eta$, and the profile is almost flat in the bulk for $\eta$ = 0.9. 
In contrast, for small $\eta = 0.5$ a 
 large  decrease of $\Omega^\prime$ to values $\ll 0.5  $ 
 (i.e., the value in the limiting  case   $\eta$ = 1,  plane Couette flow)
 is observed,  
  coinciding with a considerable 
  angular velocity gradient.

In the ultimate state of TC turbulence, the velocity boundary layer is turbulent and the profile 
 was believed to be described by the law of the wall suggested by Prandtl and von K\'arm\'an, 
\begin{equation}
U_\phi^+(\rho^+) = \kappa^{-1} \log \rho^+ + B. 
\end{equation}
Here  $\kappa$ is a 
generalized von K\'arm\'an constant and 
$\rho$ = $r-r_i$ or  $\rho$ = $r_o - r$ 
are
 the distances from the wall, presented in the usual wall units 
$\rho^+=(r-r_i)/\delta_{w,i}$ or  $\rho^+=(r_o-r )/\delta_{w,o}$, 
just as 
$U_\phi^+ = U_\phi/u_{\tau, (i,o)}$,
with 
$\delta_{w,(i,o)} = {\nu}/{u_{\tau,(i,o)}} $ and 
$u_{\tau,(i,o)} = \sqrt{{\tau_{(i,o)}}/{\rho}}$.
Since the angular velocity flux is conserved in the radial direction,
 the wall units for the outer boundary layer are directly connected to the inner ones: $u_{\tau,i} / u_{\tau,o} =1 / \eta$, and $\delta_{w,i} / \delta_{\nu,o} = \eta$. In this representation
 the von K\'arm\'an constants $\kappa$  were found to depend on $Ta$ and on
 the rotation ratio $a$,  as shown in figure~\ref{fig:profile}c taken from \cite{hou11}.

Recent work by \cite{gro14} based on the Navier-Stokes equations 
suggests that the log-law is more appropriate for the 
{\it  angular} velocity $\Omega$ rather than for the {\it azimuthal}  velocity $U_\phi$, i.e.,
\begin{equation}
\Omega^+(\rho^+) = \kappa^{-1} \text{log} \rho^+ + B,
\label{auchdasnoch}
\end{equation}
with 
$\Omega^+(\rho^+) \equiv {\left(\omega_{i,o} - \Omega(\rho^+) \right)/ \omega_{i,o}^*}$
and  $\omega_{i,o}^* = u_{i,o}/r_{i,o}$ for the inner (outer) cylinder boundary layer.  
The difference between the angular velocity and the azimuthal velocity profiles is due to the curvature. As shown in figure~\ref{fig:profile}d, for $\eta = 0.71-0.72$ 
this difference only becomes notable for large wall distances  near the middle of the TC gap. 
In contrast, for smaller radius ratio $\eta = $ 0.50, 
  the difference between the angular velocity and azimuthal velocity profiles is larger, 
   due to a relatively smaller $r_i$ as shown in  figure~\ref{fig:profile}d. 
For the largest 
$Ta = 6.2 \times 10^{12}$, the angular velocity profile at $\rho^+$ between 100 and 1000 roughly follows the Prandtl-von K\'arm\'an log-law,
with even the  same 
von K\'arm\'an constant 
 $\kappa \approx  0.4$ as found for other wall-bounded turbulent flows \citep{smi10,smi13}. 
   For lower Ta numbers the von K\'arm\'an constant $\kappa$ in eq.\ (\ref{auchdasnoch}) 
 is found to be larger than 0.4 \citep{hui13,ost14bl}, 
 presumably  since for these values
  the turbulent boundary layers are not yet fully developed and due to curvature.

\begin{figure}
\begin{center}
\includegraphics[width=1\textwidth]{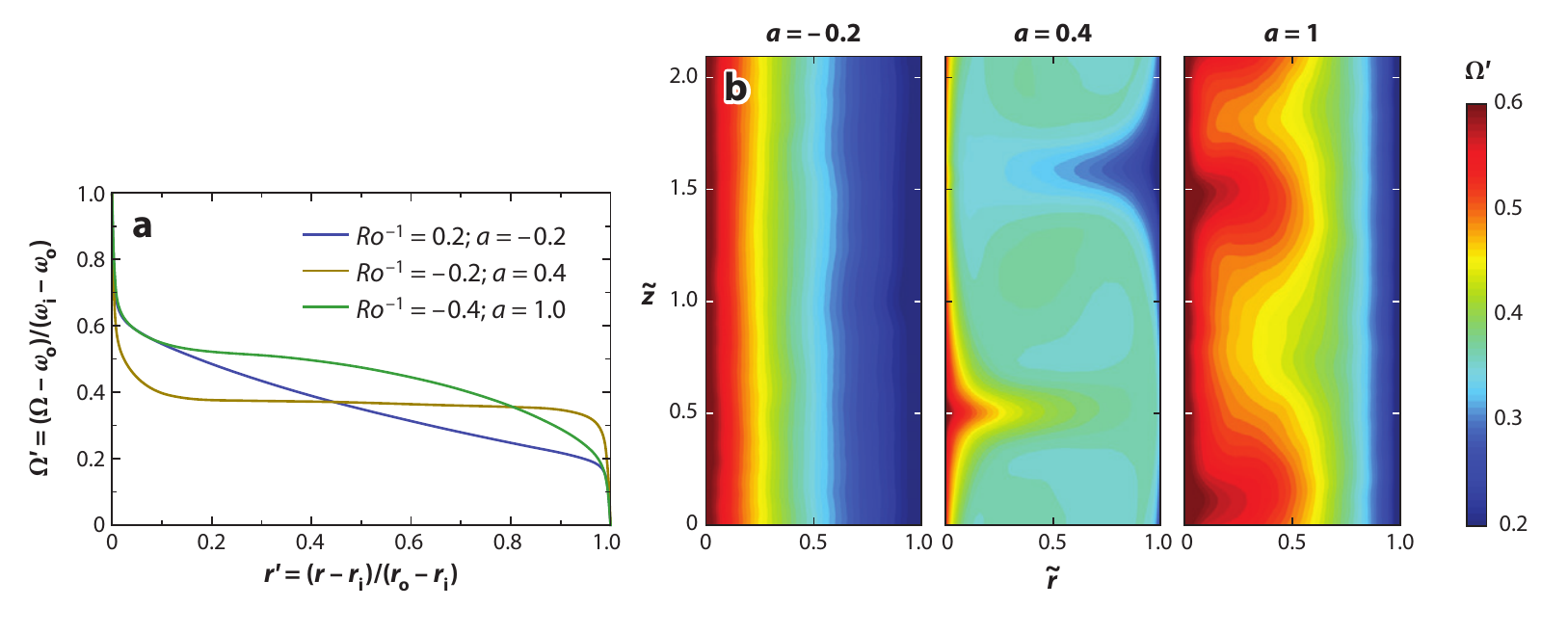}
  \caption{
  (a) Axially averaged angular velocity profiles at three values of $Ro^{-1}$ (or $a$) for $\eta =$ 0.71 and $Ta = 10^{10}$.
(b) Contour plots of the azimuthally- and time-averaged angular velocity field for $Ta = 10^{10}$, $\eta =$ 0.71 and three values of $Ro^{-1}$. The left panel corresponds to $Ro^{-1} =$ 0.2 ($a =$ -0.2) (CWCR regime) and shows no traces of axial dependence. Plumes detach rapidly into the bulk, where they  strongly mix, and thus cannot form large-scale structures. The middle panel corresponds to $Ro^{-1} = -$  0.22 ($a =$ 0.4). The reduced plume mixing enables the formation of large-scale structures, and a strong signature of them can be seen in the averaged angular velocity field. The right panel corresponds to $Ro^{-1} = -$ 0.4 ($a =$ 1)
 (SCR regime) and also shows some signatures of large-scale structures. However, these do not fully penetrate the gap but stop at the border to the Rayleigh-stable zones near the outer cylinder.
Figures taken from \cite{ost14pd}.
  }
\label{fig:profile-a}
\end{center}
\end{figure}

\subsection{Rotation ratio  dependence of profiles and optimal transport} 
We now describe in more detail 
how the profiles depend on the rotation ratio $a$ (or alternatively on the strength of the Coriolis force $\propto Ro^{-1}$). 
As seen from  figure~\ref{fig:profile-a}, 
the angular velocity profile and flow structures in the bulk 
indeed  strongly depend on this parameter, 
highlighting the effect of the Coriolis force on the  flow organization. 
 Depending on $Ro^{-1}$ (or $a$), 
  we had already defined two  regimes in section \ref{general}, namely for $Ro^{-1} > Ro_{opt}^{-1}$ 
  a co- and weakly counter-rotating regime (CWCR) 
    and for $Ro^{-1} <  Ro_{opt}^{-1}$ 
    a  strongly counter-rotating regime (SCR).  
    The transition between these two 
     regimes occurs at  the value $Ro^{-1}_{opt}$ (or $a_{opt}$)
     of the rotation ratio for which $Nu_\omega(Ro^{-1})$ is maximum.

In the CWCR regime  the Coriolis force in the Navier-Stokes equation (\ref{eq:TC_NS}) 
is balanced by
 the bulk gradient of $\Omega$, which results in a linear relationship between $Ro^{-1}$ and $\partial_r \Omega $ \citep{ost13}.
 Here $\Omega = \left<\bar{\omega} \right>_{z,t}$ is time-,  axially-,  and azimuthally-averaged 
 angular velocity. As shown with the blue line for $a = -$ 0.2
  in figure~\ref{fig:profile-a}a, the angular velocity gradient in the bulk
  is large, and to accommodate for this, there is a smaller $ \Omega $ jump across the boundary layers. In this CWCR regime, the plumes ejected from the
  boundary layers can be mixed easily in the bulk. As a consequence, the large-scale structures, which essentially consist of unmixed plumes, vanish when the driving is strong enough, see e.g. figure~\ref{fig:profile-a} at $a = -$0.2 and $Ta = 10^{10}$.  

Beyond the maximum 
the system enters   the SCR regime, i.e.,
 the outer cylinder strongly counter-rotates and generates a Coriolis force which exceeds what the $\Omega$-gradient can balance, i.e.\ $Ro^{-1} < Ro_{opt}^{-1}$ or 
  $a > a_{opt}$. The threshold value of the rotation ratio for the transition from the CWCR regime to the SCR regime  corresponds to a
 flat $\Omega$ profile \citep{gil12,ost14pd}, as shown with the red line in figure~\ref{fig:profile-a}a. This flat angular velocity profile in the bulk perfectly resembles RB turbulence for which, due to the absence of a mean temperature gradient in the bulk, the whole heat transport is conveyed by the convective term \citep{ahl09}.
Thanks to the flat angular velocity profile in the bulk, there is a large $\Omega$ jump across the boundary layers, and thus plumes detach violently from the boundary layers, strongly driving the large-scale structures. Therefore, strong large-scale structures 
form at $a = a_{opt}$, as shown in the middle panel of figure~\ref{fig:profile-a}b, and these strong mean circulations can be related to the optimum in angular velocity transport $Nu_\omega$  \citep{bra13b,ost14pd}.
\cite{ost14pd} explained also 
the radius ratio $\eta$-dependence of the angular velocity transport 
 with the flat  angular velocity profile
and the persistence of Taylor rolls: Indeed, 
$Nu_\omega (\eta )$ seems to have  its maximum close to  $\eta = 1$ (plane Couette flow), where  the angular velocity profile is flat and
the rolls are most developed. However, the detailed dependence of $Nu_\omega (\eta )$ around $\eta = 1$ still awaits a detailed numerical and experimental exploration.

In the SCR regime, the vortices cannot fully penetrate the entire domain, due to the insufficient strength of the Coriolis force. Near the outer cylinder, the flow is predominantly Rayleigh-stable. The 
angular velocity is transported mainly through intermittent turbulent bursts, instead of by 
convective transport through 
 plumes and vortices \citep{gil12,bra13b}. In this regime, large-scale coherent rolls cannot  develop
 in the entire gap,  as shown in the right panel of figure~\ref{fig:profile-a}b. Local structures are present
 only in the  zones close to the inner cylinder. The stabilization by
  the outer rotation is also clearly visible in the angular velocity profile as shown by
   the green line in figure~\ref{fig:profile-a}a, in which the outer boundary layer extends deeper into the flow, and the distinction between the bulk and the boundary layer is blurred  \citep{ost14pd}.

\begin{figure}
\begin{center}
\includegraphics[width=1\textwidth]{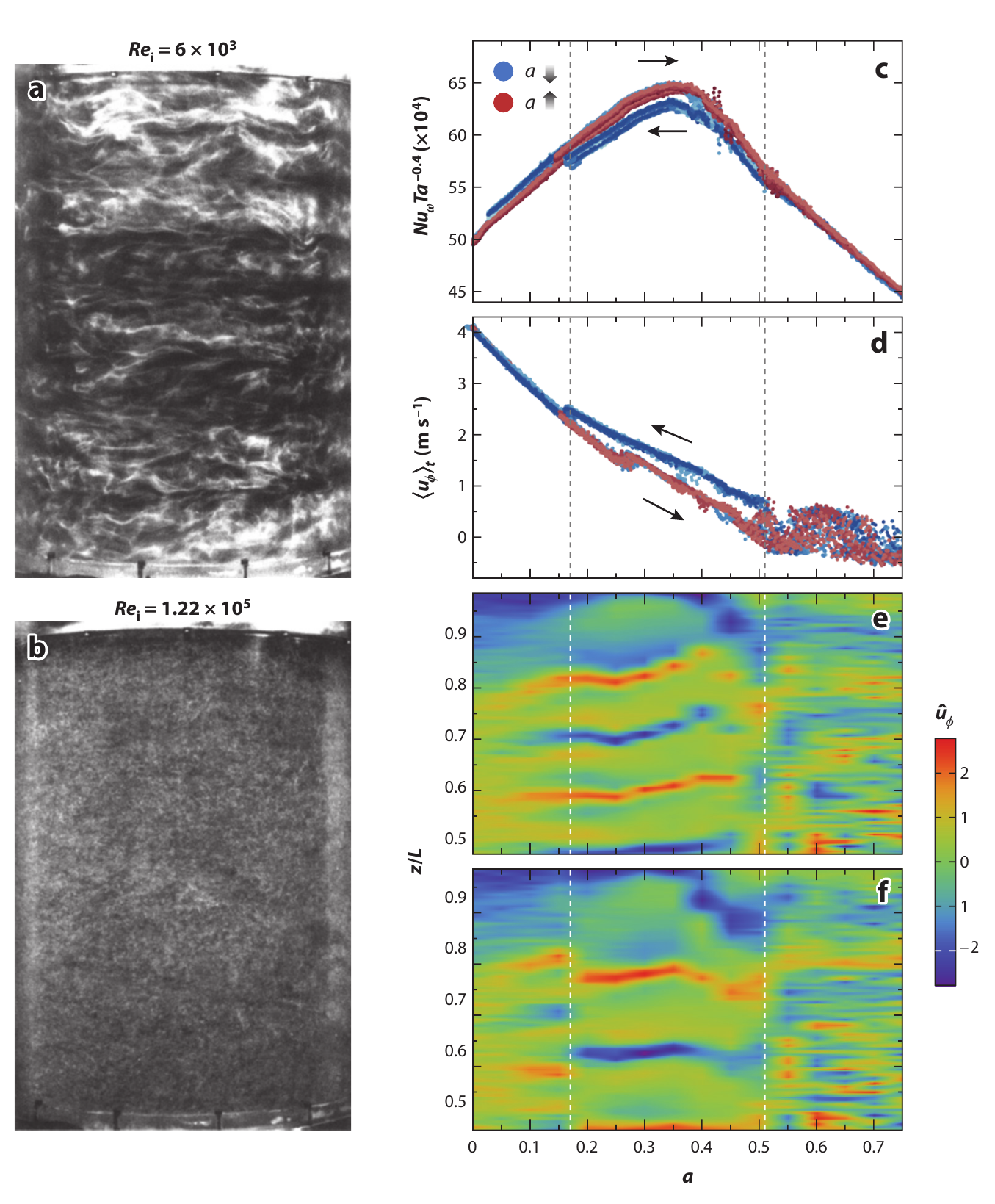}
  \caption{
  Photographs of flow states at (a) $Re_i = 6\times 10^3$ and (b) $Re_i = 1.22\times 10^5$, obtained using Kalliroscope flow visualization. The outer cylinder was stationary for these experiments. Eight vortices are visible in (a) but not in (b). Figures (a,b) taken from \cite{lat92a} with permission from the authors. 
  (c-f) Experimental results measured in the T$^3$C facility at $Ta \simeq 10^{12}$. (c) Compensated $\text{Nu}_{\omega}$ as a function of $a$. Experiments following trajectories of 
  decreasing $a$ and increasing $a$ are colored in blue and red, respectively. 
 For decreasing $a$ the flow is either in  a high or a low state for $0.17<a <0.51$, while 
 for  increasing $a$ in this regime it  always is  in the high state. 
  For $a=0.36$ the Nusselt number of the high state is $2.5\%$ larger than that of the low state.
 (d) Azimuthal velocity measured at the mid-height and mid-gap as a function of $a$ for trajectories of increasing $a$ and decreasing $a$. Same colors as in (c). For the local velocity one also sees that the system bifurcates when following 
 the trajectory of decreasing $a$ around $a=0.51$,  choosing either the high or low state for $0.17<a <0.51$.
(e) and (f) show axial scans of the normalised  angular velocity for varying $a$,
 following trajectories of increasing $a$ and decreasing $a$, respectively. One can see the presence of 4 rolls in the top half of the system in (e), while in (f) only 3 rolls are present for $0.17<a <0.51$. For $a<0.17$ and $a>0.51$ the system is in the same state, regardless of the trajectory of increasing $a$ or decreasing $a$. $\widehat{U_\phi} = (U_\phi - \left \langle U_\phi \right \rangle_z)/\sigma_a(u_\phi)$, where $\sigma_a$ is the standard deviation of $u_\phi$ for each $a$. 
Figures (c-f) taken from \cite{hui14}. 
 }
\label{fig:multi-stru}
\end{center}
\end{figure}

\subsection{Multiple turbulent state and their connection to  the  Taylor rolls}\label{multiple}

The question is whether the  turbulent structures seen in the DNS of \cite{ost14pd} up to $Re_{i,o} \sim 10^5$ and 
marked  in the phase diagram fig.\
\ref{fig:ostilla-pd}
survive in TC flows at even higher Reynolds numbers, i.e. $Re_{i,o} \gg 10^5$, 
beyond what is presently numerically possible.
And if so, what is the connection between  the turbulent structures and 
  the global angular velocity transport at very high Reynolds numbers? 

\cite{Gollub1979} observed that for increasing Reynolds number the waves on top of the Taylor vortices become increasing complex until only turbulent Taylor vortices are left. 
\cite{lew99} studied the statistics of velocity fluctuations in TC flow for $Re_i$ up to $5 \times 10^5$ 
for pure  inner cylinder rotation and found that turbulent Taylor vortices remained at their highest Reynolds number. However, the findings of \cite{lat92a} suggest that for pure inner cylinder rotation
the Taylor vortices are not present for Reynolds numbers beyond $Re_i =1.2 \times 10^5$. As shown in
 figure~\ref{fig:multi-stru}a,b, the flow structures are visible at $Re_i = 6 \times 10^3$, but  seem to 
 have disappeared at $Re_i = 1.22 \times 10^5$.

As discussed above, the flow structures in the bulk, the azimuthal velocity profile and the optimal angular velocity transport are closely connected. To correctly extrapolate to much higher Reynolds numbers it is crucial to know the characteristics of the turbulent state and the possible existence of other such states.
Kolmogorov's 1941 paradigm suggests that for strongly turbulent flows with many degrees of freedom and large fluctuations, there would only be one turbulent state as the large fluctuations would explore the entire higher dimensional phase space. 
However,  recently
 \cite{hui14} observed  conclusive evidence of {\it multiple}
  turbulent states for large Reynolds number ($Re_{i,o} \sim 10^6$, $Ta \sim 10^{12}$) TC flow in the regime of ultimate turbulence, by probing the phase space spanned by the rotation rates of the inner and outer cylinder. Furthermore, they found that the optimal transport is directly connected to the existence of  large-scale coherent structures. 

Figure~\ref{fig:multi-stru}c shows the compensated Nusselt number as a function of $a$ for \cite{hui14}'s 
 measurements with increasing $a$ (red lines) and decreasing $a$ (blue lines). As can be seen, for increasing $a$ (red lines), the torque is continuous and shows a peak around $a=0.36$, which is similar to the prior observations by \cite{gil11,pao11,gil12,mer13}. For decreasing $a$, the torque is found to be same as that with increasing $a$ at $a>0.51$ and $a<0.17$.
  However, for $0.17<a<0.51$ the torque is 
  different. For decreasing $a$ the system can enter another state around $a=0.51$ which is characterized by a lower torque (and
  thus was called `low state'), and around $a=0.17$ the system sharply jumps back to a higher torque state (`high state'). 
\cite{hui14}
 repeated these experiments to check the reproducibility. For increasing $a$, the system 
  always goes into the high state, while for decreasing $a$ it
   goes to the low state (for $0.17<a<0.51$) with a high probability (8 out of 10).

To verify that the high and low torque states originate from two different physical flow structures, \cite{hui14} measured the azimuthal velocity at half-height $z=L/2$ and at the center of the gap $r=(r_i+r_o)/2$,
 as shown in figure~\ref{fig:multi-stru}d. 
The presence of multiple states in the local measurements of the azimuthal velocity and at the same time the global measurements of the torque clearly indicates that the system can indeed be in different turbulent states, in spite
of  the very high Taylor number of $\mathcal{O}(10^{12})$ (in the ultimate regime). 

In order to further characterize the turbulent state of the system, 
\cite{hui14}  also measured the axial dependence of the azimuthal velocity in the top half of the system at varying $a$ for both the high and the low state.
 Figure~\ref{fig:multi-stru}e shows the local velocity for the high state with 5 large minima/maxima for $a\leq 0.45$, indicating that 4 turbulent Taylor vortices exist in the top half of the system. 
For $a\geq 0.5$ the state of the system is less clear, and the system appears to jump between states (without a well-defined $a$-dependence), as shown in fig.~\ref{fig:multi-stru}e. 
In the low state, figure~\ref{fig:multi-stru}f shows the same
 behavior as in the high state of fig.\ \ref{fig:multi-stru}e for $a$ outside $[0.17,0.51]$. 
 However, for $0.17<a<0.51$ it is found that the azimuthal velocity has 4 large minima/maxima in the low state, which is the signature of 3 turbulent Taylor vortices in the top half of the system.

It is surprising that multiple turbulent states exist for such large Reynolds number TC flow in the regime of ultimate turbulence. Another striking feature is that the $a$ range, 
 at which the stable (multiple) structures exist, corresponds to the $a$ range for optimal torque transport, 
 reflecting  that the optimal transport is connected to the existence of the stable large-scale coherent structures.

\section{Flow in the quasi-Keplerian regime}\label{kepler}

Astrophysical disks are ubiquitous in the universe. These so-called 
accretion disks can be of two types: 
 the proto-planetary disks and the disks around compact objects such as white dwarfs, neutron stars,  and black holes
%supermassive black holes or proto-planetary disks, and they accrete matter and thus emit radiation 
\citep{ji13}.
%in quantities which cannot be accounted for if the disks are not turbulent 
These disks are supported almost entirely by the rotation forces,
 suggesting that the rotating  matter follows a law similar to Kepler's law for planetary motion \citep{ji13}. The orbiting matter (mostly gas) has to lose its angular momentum in order to move radially inwards into the central object. 
The observed accretion rates of astrophysical disks cannot be accounted for by the transport of angular momentum resulting from pure molecular viscosity \citep{sch09,ji13}.
Therefore, 
to account for the observed transport in accretion disks, turbulence has to enter into 
 the problem. 
 Two possible instabilities have been proposed to account for this, namely a magneto-rotational instability (MRI) and a subcritical hydrodynamical instability (SHI). 
 
\begin{figure}
\begin{center}
\includegraphics[width=1\textwidth]{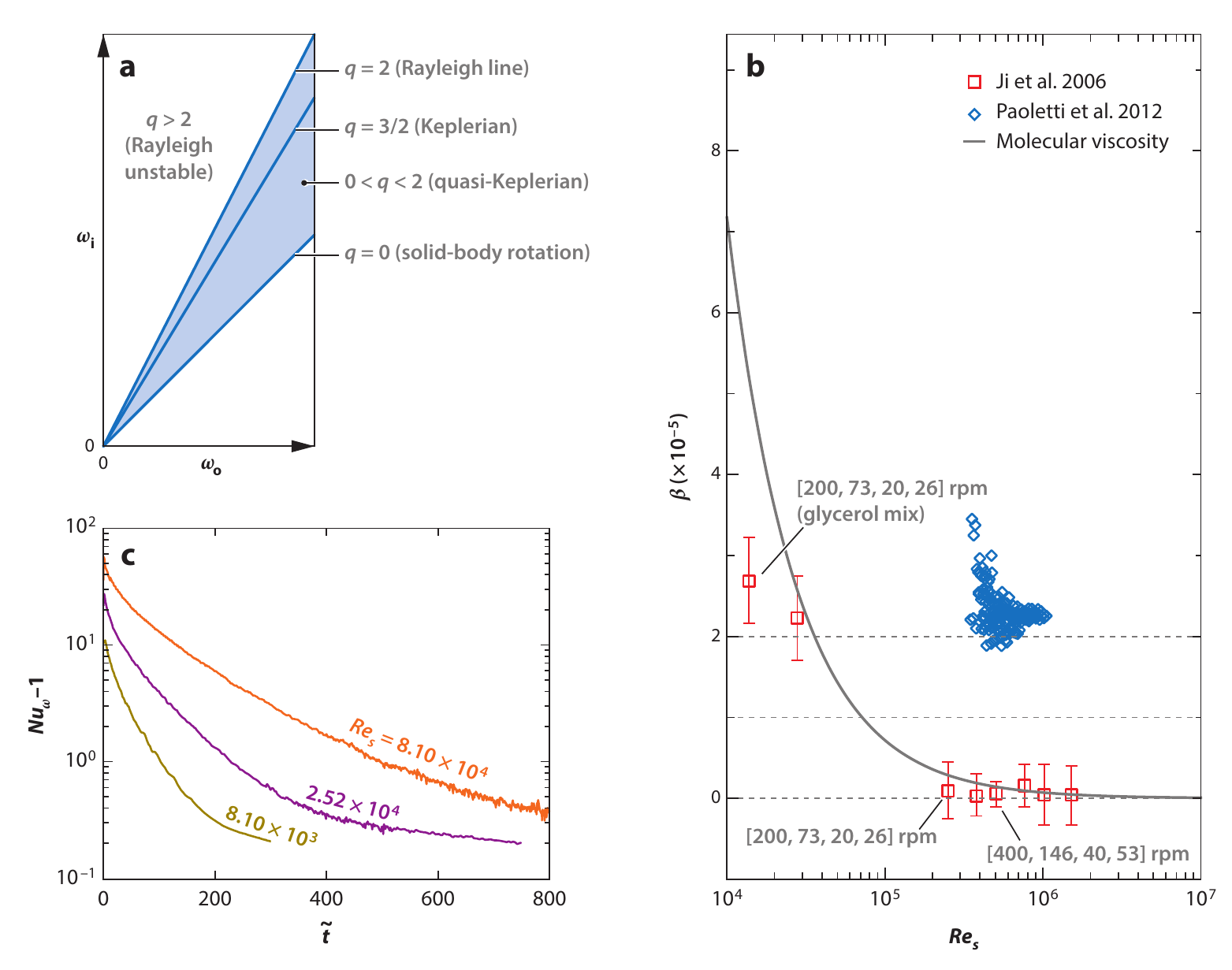}
  \caption{(a) TC parameter space in the co-rotating  regime: The different regions are shown 
  between the 
   three lines of constant $q$, characterising 
   the Rayleigh criterion ($q = 2$), the Keplerian case  ($q = 3/2$), and solid-body rotation ($q = 0$).  
   Figure taken from \cite{nor15}. 
  (b) Dimensionless turbulent viscosity $\beta$ vs.\ the shear Reynolds number $Re_s$.  
   The black line is the molecular viscous value: 
  $\beta_{vic} =\nu/[\bar{r}^3(\omega_i-\omega_o)/d] $ $=\frac{16\eta(1-\eta)^2}{(1+\eta)^4}Re_s^{-1}$, where $\bar{r} = (r_o+r_i)/2$. Figure adapted from \cite{ji06} and \cite{pao12}.  
    (c) Semi-log time series of $Nu_\omega - 1$ at the inner cylinder for $Ro^{-1}$ = 1.22 
   (implying $q$ = 1.5)
  and three values of the shear Reynolds number, namely  
  $Re_s = 8.10 \times 10^3$ (black line), 
    $Re_s = 2.52 \times 10^4$ (blue dashed line), and  $Re_s = 8.10 \times 10^4$ (red dashed-dotted line).  Figure taken from \cite{ost14decay}. 
    }
\label{fig:kepler}
\end{center}
\end{figure}

 The latter one is the reason for which  the TC system is relevant in this context, as it has been 
suggested  as model system for studying the transport behavior in accretion disks \citep[see e.g.][]{richard_thesis_2001,dub05,avi12,ji13}.
The astrophysical community uses 
different control and response parameters from 
 those introduced in section \ref{param}. 
More specifically, 
rather than using the Taylor number Ta (or $Re_{i,o}$) and 
the rotation ratio $a$ (or the inverse Rossby number $Ro^{-1}$)
 as control parameter, that community uses a shear  Reynolds number $Re_s$ and 
the  parameter $q$, which are defined through the relations \citep{pao12,ji06}
\begin{equation}
Re_s =  {2 \over 1 + \eta } ~ | \eta Re_o - Re_i |, \qquad
\omega_i/\omega_o = \eta^{-q}.
\end{equation} 
The $q$ parameter is real for the 
co-rotating situation, which is the only case we consider 
 in this section. The parameter space at different $q$ values is shown in figure~\ref{fig:kepler}a.
%Solid body rotation $\omega_i = \omega_o$ corresponds to $q = 0$, $\omega_i > \omega_o$ means $q > 0$, $\omega_i < \omega_o$ gives $q < 0$, and the pure inner and pure outer rotation correspond to $q = + \infty$ and  $q = - \infty$, respectively. 
 TC flow with  $q > 2$ (discussed in the previous sections) is  linearly unstable at sufficiently high Reynolds numbers,
  due to 
  the centrifugal instability. 
  In the regime  $q < 2$,  in 
   which the angular momentum $\Omega r^2$ radially increases with increasing $r$ ($\omega_i r_i^2 < \omega_o r_o^2$), the flow is linearly stable according to the Rayleigh criterion \citep{ray17}. 
This 
 includes subrotation ($\omega_i < \omega_o$), solid-body ($\omega_i = \omega_o$), and super-rotation ($\omega_i > \omega_o$).
The flow in  the region ($\omega_i > \omega_o$ and $\omega_i r_i^2 < \omega_o r_o^2$, i.e.,  $0<q<2$) is often referred to as  {\it quasi-Keplerian flow}, since it includes cylinder rotation rates ($q = 3/2$) obeying Kepler's law relating orbital radius and period.
As response parameter the astrophysical community uses
the dimensionless turbulent viscosity $\beta$ \citep{ric99,dub05,her05} defined through
\begin{equation}
\beta = {2 (1-\eta)^4  \over \pi \eta^{2}} ~
\frac{G}{Re_s^2} = {2 (1-\eta)^4  \over \pi \eta^{2}} ~
\frac{Nu_\omega G_{lam}}{Re_s^2}  
,
\end{equation}
rather than the dimensionless torque G or the angular velocity $Nu_\omega$.

The key question now is whether the TC flow in the linearly stable quasi-Keplerian regime ($0<q<2$) 
can nevertheless become  unstable to finite size perturbations, due to a subcritical transition to turbulence. 
Indeed,  in spite of linear stability, at large enough Reynolds numbers
shear flow {\it may} get unstable, namely through the so-called non-normal - nonlinear mechanism,
as well known from pipe flow or plane Couette flow, see e.g.\
\cite{tre93,grossmann2000,eck07}.

 \cite{pao11} and \cite{pao12} examined the flow stability
  by measuring the global torque as a function of Reynolds number using the Maryland TC facility
  ($\eta = 0.7245$, aspect ratio $\Gamma = 11.47$, measurements of the torque only with the central
   part of the inner cylinder)
   in the quasi-Keplerian regime. 
The axial boundaries (end plates) in their experiments rotated with the outer cylinder. The measured dimensionless turbulent viscosity $\beta$ as a  
function of $Re_s$  is  shown in figure~\ref{fig:kepler}b. 
%As shown in the figure, the dimensionless 
%turbulent viscosity $\beta$ 
At $Re_s \simeq 7 \times 10^5$ it is  $\beta \approx 2 \times 10^{-5}$, which is much larger than the corresponding value of the laminar viscous transport, thus suggesting turbulent  behavior. 
Extrapolating towards astrophysical Reynolds number ($\sim 10^{13}$) by using the formula of \cite{her05}, \cite{pao12} estimated 
 that then $\beta \approx 7.5 \times 10^{-6}$, which is consistent with the values observed in disks around T Tauri stars ($7\times 10^{-8}$ to $3.5 \times 10^{-5}$).
However, by means of Laser Doppler Anemometry (LDA) of the 
 azimuthal velocity profiles in the quasi-Keplerian regime in the T$^3$C facility, 
  \cite{nor15} 
  found  that the angular momentum is transported axially to the axial boundaries, suggesting that TC 
  flow with end plates attached to the outer cylinder is an imperfect model for accretion disk flows.

\cite{ji06} studied the angular momentum transport in TC flow in the quasi-Keplerian regime
with a different experimental setup.
 The  radius ratio and aspect ratio were $\eta =  0.35$  and $\Gamma = 2.1$, respectively. 
 In order to minimize Ekman circulation, the end plates were split into two rings that could independently rotate (with the rotation rates of $\omega_3$ and $\omega_4$) with respect to the inner and outer cylinders. By carefully controlling the rotation speeds [$\omega_i$, $\omega_3$, $\omega_4$, $\omega_o$], the azimuthal velocity profiles in the gap were tuned to match the quasi-Keplerian flow profiles, $\Omega(r) = r^q$ with $0<q<2$. The angular momentum transport was quantified by measuring the
 local angular velocity flux, $\rho r \left<u_\theta' u_r'\right>_t$, where $u_\theta'$ and $u_r'$ were the instantaneous velocity fluctuations in radial and azimuthal directions, and $\left< \right>_t$ stands
  for averaging over time. 
  The experimental  results for $\beta$ versus $Re_s$ are shown as
   open squares in figure~\ref{fig:kepler}b.  When the end-ring plates were
    optimized to produce Couette profiles, the values of $\beta$ were
     found to be around $0.72 \times 10^{-6}$ with a standard deviation of $2.7 \times 10^{-6}$ at 
     $Re_s$ 
     from $3 \times 10^5$ to $2 \times 10^6$. These values are indistinguishable from the corresponding molecular viscous transport as shown with the solid line in the figure.  The work by \cite{ji06} thus
     demonstrates
      that the axial boundaries can profoundly influence linearly stable flows
      and purely hydrodynamic quasi-Keplerian flows, under proper boundary conditions and at large enough Reynolds numbers,
      but  cannot transport angular momentum at astrophysically relevant rates.

To settle the issue of  end plate effects, 
\citet{avi12} performed direct numerical simulations using both Maryland \citep{pao11} and Princeton \citep{ji06} TC geometries at $Re_s$ up to $6.5 \times 10^3$.  He found that the  
end ring plates   do drive secondary flows that enhance the global angular momentum transport in both the Maryland and the Princeton geometry.
His studies thus indicate  that the current laboratory TC apparatuses designed to approximate flow profiles of accretion disks suffer from the imposed boundary conditions (end plates and finite aspect ratio).  
However,  one would also expect that with increasing $Re_s$ the effect of the end plates
will get less, as  suggested by figure \ref{fig:kepler}b. 

To completely prevent end plate effects on the transition, \cite{ost14decay} investigated this issue numerically, 
 using periodic boundary conditions in the axial direction at high Reynolds numbers. The shear Reynolds number they achieved in their direct numerical simulations (DNS) was up to $10^5$ with a radius ratio of 0.5. The procedures they used for the simulations were as follows:
  The simulations started with a turbulent field, corresponding to a pure inner cylinder rotation in the laboratory frame
   ($a = 0$, $Ro^{-1} = 0$). Then the outer cylinder was switched on such that  
   the rotation ratio was in the quasi-Keplerian regime. As shown in figure~\ref{fig:kepler}c, the Nusselt number decreased as a function of time down to a value corresponding to purely non-vortical laminar flow. 
This work thus showed  not only that the TC system at $0<q<2$ is linearly stable in this geometry, but 
that even an initially turbulent flow with shear large Reynolds numbers up to $10^5$ 
decays towards the linearly stable regime, due to stabilizing counter-rotation.  This finding is consistent with the experimental results by \cite{edl14}. 
Figure~\ref{fig:kepler}c also shows that
the decay time for turbulence becomes longer with increasing shear Reynolds number. 

In summary, the present evidence \citep{ji06,ost14decay,edl14}  indicates that the TC system in the quasi-Keplerian regime is linearly stable at 
shear Reynolds number up to $Re_s \sim 10^5 - 10^6$. 
However, one  cannot exclude 
 subcritical transitions to turbulence in the quasi-Keplerian regime for TC flow at much higher shear Reynolds numbers
and  it is thus necessary to examine  the TC system  at much higher shear Reynolds numbers. 

We also note that  for $q< 0$, i.e., linearly stable but outside the quasi-Keplerin regime, 
 subcritical transitions to turbulence may occur: E.g.\ \cite{bor10} studied the decay characteristics in TC flow for pure outer cylinder rotation at intermediate Reynolds numbers (up to $Re_o \sim 10^4$) and 
 found a  subcritical transition to turbulence. 
Also \cite{bur12} performed experiments for pure outer cylinder rotation, finding subcritical transitions to spiral or intermittent
 turbulence, too, which strongly depended
on the radius ratio $\eta$ and the chosen end-cap configurations.

\section{Summary and outlook}\label{open}

\noindent 
Summary points: 
\begin{itemize}
\item The last decade has witnessed a tremendous extension of the experimentally and numerically
studied TC parameter space towards 
the strongly turbulent regime, including co- and counter-rotation of the cylinders. 
\item 
In the {\it classical regime} the bulk is turbulent and  
the BLs are of Prandtl-Blasius (laminar) type, whereas 
 in the {\it ultimate regime} also  the BLs have  become turbulent, i.e., are of Prandtl-von K\'arm\'an type.  
 The transition towards the ultimate regime occurs  around $Ta^* \simeq  3 \times 10^8$ (for $\eta = 0.71$). 
 Due to the more efficient mechanical driving  
 this is much earlier than  in the
 analogous, but thermally driven  RB system, where the transition towards the ultimate regime occurs at $Ra^* \simeq 10^{14}$.   
\item In the classical regime, for pure  inner cylinder rotation 
the effective scaling exponent $\gamma$ of $Nu_\omega \sim Ta^\gamma$ is less than 1/3, whereas in the ultimate regime, 
for any rotation ratio $\gamma$ is larger than 1/3 and follows the theoretical prediction of $Nu_\omega \sim Ta^{1/2} \times$ log-corrections.
\item At large enough Taylor number, $Nu_\omega$ has a non-monotonic dependence on 
the (negative)  rotation ratio $a$ with a pronounced maximum at $a_{opt} >   0$, at
 which the angular velocity transport is optimal. The value 
  $a_{opt}$ depends on $\eta$: It increases with increasing $\eta$ for $\eta \lesssim 0.8$ and 
   seems to saturate at larger  $\eta$. 
\item For optimal $Nu_\omega $ the angular velocity profile is flat and the flow displays 
pronounced Taylor rolls, in spite of the strong turbulence. The roll structure allows for the existence of multiple turbulent states
with different transport and flow properties. 
\item For large enough Ta, the profiles of the BLs in  the ultimate regime are well described by a log-law 
for the angular velocity,  $\Omega^+(\rho^+) = \kappa^{-1} \log  \rho^+ + B$, with the well-known von K\'arm\'an 
constant $\kappa \approx  0.4$.
\item The current laboratory TC apparatuses designed to approximate flow profiles of accretion disks suffer from the imposed boundary conditions in axial direction. The present experiments and numerical simulations
 suggest that the TC system in the quasi-Keplerian regime is linearly stable for 
  shear Reynolds number up to $Re_s \sim 10^5 - 10^6$ and presumably beyond, where hitherto no experiments exist.
  But note that the shear Reynolds numbers relevant in astrophysical circumstellar disks are as large as 
  $10^{13}$ and higher \citep{her05}.  
\end{itemize}

\bigskip 

\noindent 
Future issues:
\begin{itemize}
\item A rigorous
 theoretical understanding of the dependence of $a_{opt}$ on $\eta$ is still missing. 
In particular, for $\eta \rightarrow 1 $ different mechanisms may be at play as compared to the situation at 
smaller $\eta$ on which
we focused here. 
\item  In the other limiting case at very small radius ratios $\eta \rightarrow 0$ future investigations 
are needed, helping to quantitatively examine the curvature effects of the cylinders. One should also try to understand
why   the transitional  Taylor number $Ta^*$ to the ultimate regime  becomes larger at  smaller radius ratios. 
\item When further increasing Ta to $Ta \gg 10^{13}$: Will  the large-scale coherent structures (turbulent Taylor rolls)
continue to exist? If so, will  multiple turbulent states still coexist or will the fluctuation be so large  that
the turbulent dynamics meanders between these states? 
\item The review did not touch upon the statistics of the turbulent fluctuations, structure functions, and spectra. 
\cite{lat92a} and \cite{hui13pre} experimentally showed that these are different as compared to homogeneous isotropic turbulence, but the reason is unknown. 
\item As the TC system is closed and can be very well controlled,
it  is excellently  suited to  study  {\it multiphase}
 turbulent flows, i.e., 
 flows with particles, droplets, bubbles \citep{ber05,ber07,gil13}, 
 and even with vapor bubbles nucleating close to  phase transitions. 
\item For the same reasons the TC system is very well suited to study the interaction of turbulence with rough walls \citep{cad97,ber03},
with superhydrophobic wall \citep{sri15}, 
and with 
 micro- and nano-structured walls, where the structures are either of geometric or  of chemical nature.
\end{itemize}

\section*{Disclosure statement}
The authors are not aware of any biases that might be perceived as affecting the objectivity of this review.

%\vspace{-0.4cm}

\section*{Acknowledgements}
We thank all our coworkers and colleagues
for their contributions to our understanding 
of this great problem, for the many stimulating 
discussions we had the privilege to
enjoy with them over  the years, and for their valuable comments on this 
manuscript.  -- 
We also gratefully acknowledge FOM, STW, NWO, and ERC (via an Advanced Grant) for financial support over the years.

%\input{arfm-bbl.tex}

%\bibliographystyle{ar-style1}
%\bibliography{literatur}

\end{document}